\documentclass[twocolumn]{emulateapj}

\shorttitle{Disk assembly - TF Relation to $z$$\sim$1.3}
\shortauthors{Miller et al.}
\journalinfo{\@Accepted for publication in the Astrophysical Journal}

\begin{document}
\title{
The Assembly History of Disk Galaxies: \\ I -- 
The Tully-Fisher Relation to $z\simeq$ 1.3 from Deep Exposures with DEIMOS
    }

\author{
Sarah H. Miller\altaffilmark{1,3},
Kevin Bundy\altaffilmark{2},
Mark Sullivan\altaffilmark{1},
Richard S. Ellis\altaffilmark{3}, \&
Tommaso Treu\altaffilmark{4}
}

\email{s.miller1@physics.ox.ac.uk}

\altaffiltext{1}{Oxford Astrophysics, Oxford, OX1 3RH, UK}
\altaffiltext{2}{Astronomy Department, University of California, Berkeley, CA 94720}
\altaffiltext{3}{California Institute of Technology, Pasadena, CA 91125}
\altaffiltext{4}{UC Santa Barbara Physics, Santa Barbara, CA 93106}

\begin{abstract}

We present new measures of the evolving scaling relations between stellar mass,
luminosity and rotational velocity for a morphologically-inclusive sample of 129
disk-like galaxies with $z_{AB}<$~22.5 in the redshift range 0.2 ~$<z<$~1.3, based on
spectra from DEIMOS on the Keck II telescope, multi-color HST ACS photometry,
and ground-based Ks-band imaging. A unique feature of our survey is the extended
spectroscopic integration times, leading to significant improvements in
determining characteristic rotational velocities for each galaxy. Rotation curves are
reliably traced to the radius where they begin to flatten for $\sim$90\% of our sample,
and we model the HST-resolved bulge and disk components in order to accurately
de-project our measured velocities, accounting for seeing and dispersion. 
We demonstrate the merit of these advances by recovering an intrinsic
scatter on the stellar mass Tully-Fisher relation a factor of 2-3 less than in previous
studies at intermediate redshift and comparable to that of locally-determined
relations. With our increased precision, we find the relation is well-established by
$\langle z\rangle \sim1$, with no significant evolution to $\langle z\rangle \sim0.3$,
$\Delta M_{\ast} \sim 0.04 \pm 0.07$ dex. 
A clearer trend of evolution is seen in the B-band Tully-Fisher 
relation corresponding to a decline in luminosity of $\Delta M_{B} \sim 0.85 \pm 0.28$ 
magnitudes at fixed velocity over the same redshift range, reflecting the changes 
in star formation over this period. As an illustration of the opportunities possible 
when gas masses are available for a sample such as ours, we show how our 
dynamical and stellar mass data can be used to evaluate the likely contributions 
of baryons and dark matter to the assembly history of spiral galaxies.
  
\end{abstract}

\section{Introduction}

A major challenge for $\Lambda$CDM structure formation lies in understanding how the baryonic components of galaxies assemble within dark matter halos. Although baryons represent only one-sixth of the gravitating matter in WMAP cosmology \citep{spergel07,seljak05}, their dissipative properties suggest they dominate the inner regions of luminous galaxies \citep{blumen1986}. Determining the interplay between dark matter and baryons is critical for predicting the evolution of density profiles, substructure, shapes, and angular momentum of galaxies \citep{govern2007, shlosm2009}. One of the most significant challenges is reproducing the detailed characteristics of rotationally supported disk galaxies which represent the dominant fraction of present-day luminous systems \citep{ellis09}.

Observational efforts in this challenge have focused on the Tully-Fisher (TF) relation \citep{tully1977} and its past evolution. This important scaling relation, which correlates disk luminosity with rotational velocity, provides an essential benchmark for verifying theoretical models based on the standard dark matter picture. Early N-body simulations as well as semi-analytic models produced galaxies that rotate too fast at a given luminosity \citep{vanden2000, momao2000, ekenav2001, benson2003, dutton2007}.  Caused by a transfer of angular momentum from baryons to the dark matter halo, this deficiency has since been mitigated by improved resolution, as well as the introduction of feedback \citep{steinm1999}, e.g. from supernovae \citep{govern2007, pionte2009}. However reproducing the absolute values observed in the scaling relation has remained problematic. 

Despite these challenges, the theoretical understanding of disk galaxy scaling relations and their evolution has made some improvement over the past decades.  Using an adjustment to the rotational velocity derived from their hydrodynamic simulations to account for over-merging, \citet{portin2007} were able to match the observed local TF relation, and claim a modest evolution to $z\sim1$. However the predictive power is tempered by an unknown dependence on redshift of this adjustment. Semi-analytic models have also worked to match observations and provide further insight on the physical interpretation of evolution in the TF relation. Some controversy remains over whether the central regions of galaxy halos are subject to adiabatic contraction \citep{somerv2008a}, broadly maintain a non-evolving density profile \citep{wechsl2002}, or permit adiabatic expansion  \citep{dutton2011b}. Regardless of the exact evolutionary response of the inner halo, the persistent picture is one in which the baryonic component grows in tandem with the dynamical mass \citep{fallefstathiou1980,dalcanton1997,momaowhite1998}. Gas may cool from the halo or from externally-sourced streams, increasing the disk scale length as stars form.  In this framework, while any given galaxy is predicted to grow by factors of 1.2-2 in stellar mass, dynamical mass, scale radius, and luminosity since $z \sim 1$ (modulo evolutionary corrections), this growth typically occurs {\em along} scaling relations, reducing the evolutionary signals accessible to observations. 

Observational progress in testing these pictures of disk assembly has been similarly slow. There are significant technical challenges in making the necessary measurements at intermediate redshift and, as a result, there are discrepant conclusions with regard to evolutionary trends in the literature.  In part, this may reflect different ways in which intermediate redshift disk galaxies are selected. \citet{vogt1996, vogt1997} undertook an important pioneering study, finding a modest increase in luminosity ($\Delta M_{B} \sim 0.6$) at fixed velocity to $z\sim1$, but deduced this represented only an upper limit to possible evolution because of sample biases and other assumptions. Subsequent optical-based studies have presented mixed conclusions. A key uncertainty is whether to address evolution in the overall mass-to-light ratio independent of luminosity (i.e. a zero-point shift with redshift) as discussed by \citet{rix2007}, \citet{bamfor2006}, and \citet{fernan2009,fernan2010}, or whether to permit luminosity dependent evolution (i.e. changes in the TF slope) as discussed by \citet{ziegle2002} and \citet{bohm2004}.  TF studies at infrared wavelengths are less affected by biases induced by short-term star formation activity and early surveys found no convincing evolution \citep{consel2005, flores2006}. However, by contrast, \citet{puech2008} claim from near-IR measures that disks were overall {\it less luminous} in the past. Clearly the rest wavelength at which the luminosity is sampled is a key parameter: \citet{fernan2010} claim evolution in the $B$-band but none in redder bands, while \citet{weiner2006b, weiner2006a} find  evolution in the slope of the infrared TF relation consistent with that seen in the blue relation, however they observe little evolution in infrared zero-point. \citet{moran2007} have emphasized the importance of environmental influences which can be deduced by considering the scatter in the TF relation as a function of local density.
 
In view of this, a more physically-relevant approach for understanding the assembly history of disks may be to consider the  
{\em stellar mass TF relation} ($M_{\ast}$-TF)  which, notwithstanding the difficulty of estimating gas fractions, provides the most robust route to understanding the interplay between baryons and dark matter in disk galaxies.  Stellar masses are derived using population model fits to multi-color photometry for galaxies of known redshift, assuming an initial stellar mass function \citep{brinch2000, bundy2005}.  Although the low redshift $M_{\ast}$-TF relation is well-constrained \citep{bellde2001, pizagn2005, meyer2008}, those at intermediate redshift \citep{consel2005, flores2006, atkins2007} reveal a larger scatter than seen in the traditional TF relations, suggestive of additional uncertainties.  A recurrent topic of discussion in the literature is sample selection and whether evolution seen in both the TF relation and its scatter is driven by redshift-dependent selection criteria. The inclusion of more early types and dynamically disturbed galaxies likely broadens the intrinsic scatter. In an attempt to include more kinematically-disturbed galaxies, often excluded without good cause in TF studies, \citet{weiner2006a} and \citet{kassin2007} have included an additional kinematic term, $S_{0.5}$, which accommodates the isotropic velocity width of the observed emission lines and reduces the scatter from their classic $M_{\ast}$-rotational velocity relation. \citet{kassin2007} detect no significant evolution in the TF relation, including the $S_{0.5}$ relation, over $0.1 < z < 1.2$. However, combining a measure of the velocity dispersion with the disk angular momentum may obscure information about the rotational support of the disk.

Using integral field unit (IFU) spectrographs, \citet{flores2006} and \citet{puech2008} have produced intensity, velocity, and dispersion maps of galaxies at intermediate redshifts that demonstrate the unique advantage of a second spatial dimension in modeling the velocity field and accounting for projection effects. So far, the IFU-based samples are fairly modest in size and sample brighter sources compared to those reached with multi-slit techniques. Moreoever, the spaxel resolution is often lower than for the highest-resolution slit spectroscopy.  As we will show in this paper, the spectroscopic signal-to-noise is an equally important factor in making progress because it determines the radial extent to which emission lines can be traced and as a result, whether the adopted velocity measure requires extrapolation.

Both IFUs and slit spectroscopy have been successfully employed at $z \approx 2-3$, where near-IR studies can take advantage of redshifted H$\alpha$ lines---the brightest kinematic tracer---as well as improved spatial resolution from adaptive optics.  Results at these redshifts may indicate the emergence of regular scaling relations from more complex and disordered dynamical states.  With adequate sampling, at least one-third of $z\simeq$ 2-3 star-forming galaxies show ordered rotation \citep{shapir2008,stark2008,jones2010} and reveal significantly higher velocity dispersions than local counterparts (e.g., \citet{genzel2006}).  \citet{cresci2009} have measured the stellar mass TF relation at $z \approx 2$ using SINFONI observations of 18 rotation-dominated systems in the SINS survey \citep{forste2009}. The slope of their measured relation is consistent with that seen in local observations but offset towards lower $M_*$ at fixed velocity by $\sim$0.5 dex.  While necessarily biased toward massive systems with well-ordered rotation, these may be representative of gas rich systems in transition to z$\sim$1 disks \citep{tacconi2010}. \citet{gneruc2010} have recently measured the TF relation at $z\sim3$ from SINFONI IFU data, but because of the large scatter observed, they suggest the TF relation has not yet been established. However all points on the relation are consistent with the favored vector of disk assembly theory, with a lower average stellar-to-dynamical mass ratio than found in the local universe.

The present survey was motivated by our desire to chart and understand this evolution from a prevalence of disturbed and complex dynamical states observed at high $z$ to the well-ordered rotation of local spirals. To make progress, we seek to determine the stellar mass TF relation over the redshift range 0.2~$<z<$~1.3 with spectroscopic exposures 3 to 8 times that of previous studies.  By including disk systems selected from HST ACS data with irregular or distorted morphologies, we hope to avoid biases based on selecting mature, well-ordered disks \citep{vogt1996,vogt1997}.  Our study not only allows us to chart evolution in a large sample down to fainter limits and masses than is possible at $z \sim 2$, but the improved precision we demonstrate enables us to measure a robust TF relation only a few Gyr later. The scatter we observe should provide a valuable indication of the rate at which disks settle onto the local TF relation. To fully utilize the gains in signal-to-noise ratio from long exposures, we develop an improved method for extracting the rotation curves of galaxies at intermediate redshift via a new modeling code. While our results are based on slit spectroscopy, unlike most previous slit-based studies \citep{weiner2006b, kassin2007, bohm2007}, we are able to align the spectral slits on our masks with the HST-measured major axis, significantly improving the fidelity of our recovered rotation curves at $z \sim 1$. We aim to avoid introducing a bias towards aligning the often more extended and brighter objects at lower redshift over the higher redshift objects observed at smaller angular scale, since doing so could introduce an extraneous evolution in the offset and scatter of the Tully-Fisher relation with redshift.

A plan of the paper follows. In \S 2, we describe our sample, the Keck spectroscopic data and the HST ACS resolved photometry in the north and south GOODS fields. Noting the limitations of earlier work, \S 3 introduces a new procedure for the analysis of rotation curve data. We justify our chosen method, discuss error estimation, and compare to previous work. In \S 4 we present the various TF relations, and in \S 5 we discuss methods for deriving dynamical masses to compare to baryonic mass estimates for a physical interpretation of our results. Finally, \S 6 summarizes the overall conclusions for the assembly history of disk galaxies.

Throughout the paper we adopt a \citet{chabri2003} initial mass function and a $\Omega_{\Lambda}$ = 0.7, $\Omega_{m}$ = 0.3, $H_0$ = 70 km sec$^{-1}$ Mpc$^{-1}$
cosmology. All magnitudes refer to those in the AB system.

\section{Data}

\begin{figure*}
\begin{center}
\includegraphics[width=4.3in]{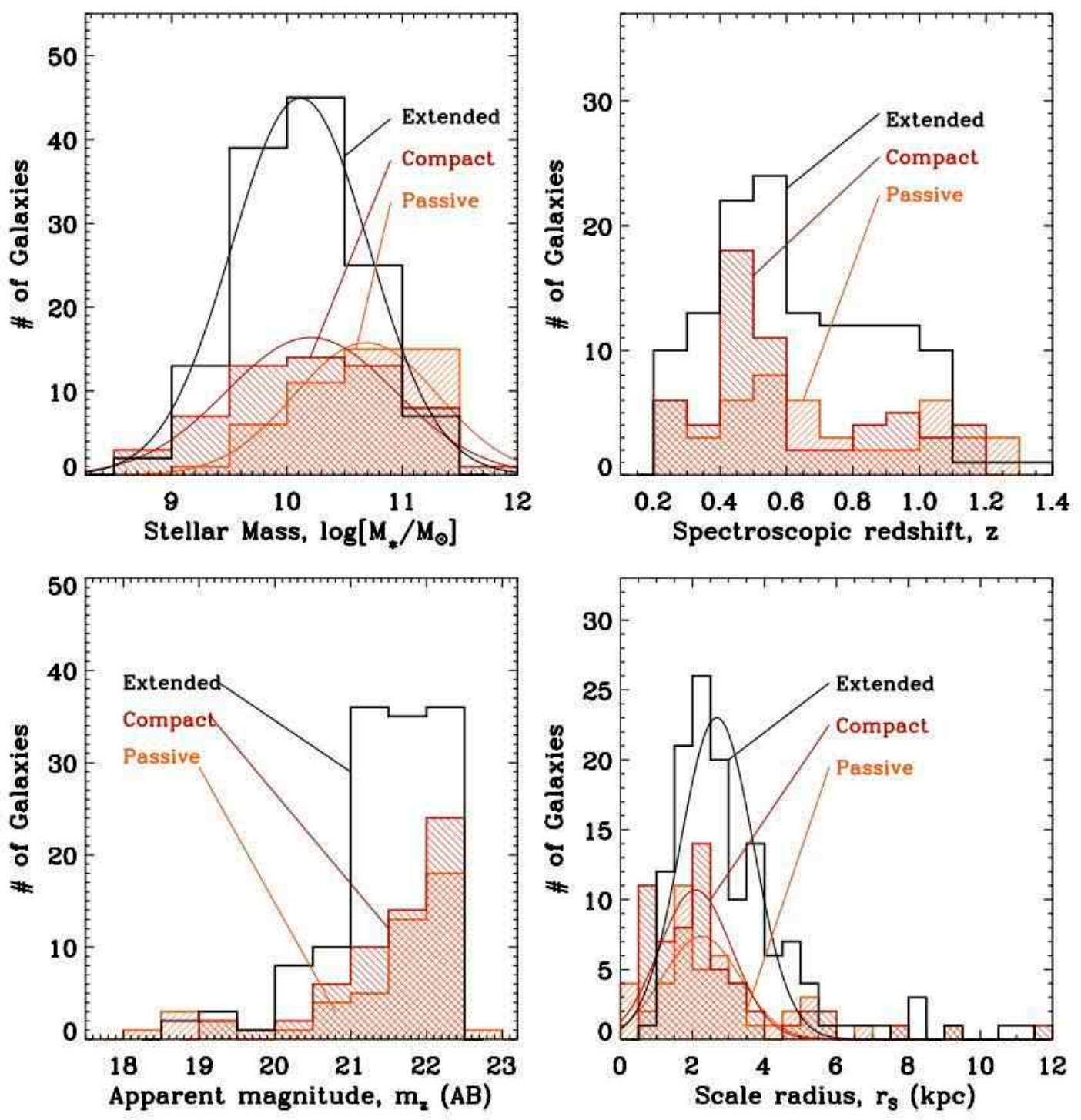}
\caption{The properties of galaxies in our sample in terms of the distributions of stellar mass
  estimates, spectroscopic redshifts, $z_{F850LP}$ apparent magnitudes, and disk scale radii. Each
  histogram is partitioned according to galaxies which have extended emission lines (129),  passive
  spectra with no emission (49) and spectrally compact sources where emission does not extend beyond the
  central-most regions of their disks (59) (see \S \ref{sec:spec} and Table \ref{table_sample} for details).}
\label{fig:pasmass_hist}
\end{center}
\end{figure*}

A prerequisite for constructing a disk galaxy sample suitable for measuring the evolution of the Tully-Fisher relation is deep HST imaging essential for morphological selection, resolved photometry and accurate disk position angle data for the multi-slit spectroscopic campaign. Both northern and southern GOODS fields \citep{dickin2003} are visible from the Keck observatory and remain the most appropriate areas for such a study given the unique availability of deep multi-color ACS data. In this section we introduce our sample selection criteria (\S \ref{sec:sample}) and the spectroscopic data used for measuring the internal dynamics of our sample (\S \ref{sec:spec}). We also introduce the photometric data used for measuring galaxy sizes and shapes (\S \ref{sec:phot}) and stellar mass estimates from SED fitting (\S \ref{sec:masses}).

\subsection{Sample Selection}\label{sec:sample}

Our goal in the morphological selection of disk targets is to be inclusive of all galaxies with disk-like structure, avoiding the temptation of selecting the most relaxed and ``well-behaved'' spirals in favor of a more complete census, including the more disturbed and morphologically abnormal population.  Sources were selected visually by co-author RSE from a $z_ {F850LP} < 22.5$ sample of 2978 galaxies, in the GOODS North and South fields, discussed by \citet{bundy2005}. A key difference with earlier work (i.e. \citet{vogt1996,vogt1997,consel2005}) is the inclusion of less mature morphological types which contain some evidence of disk-like structure, as well as systems that may be interacting. We included visually irregular systems with elongated features, galaxies with asymmetric and clumpy light distributions. We also included disks with dominant bulges. The main aim of broadening the morphological selection criteria was to avoid potential biases associated with selecting only symmetric spirals which may represent the end point of evolution and consequently bias us to locating mature systems.  Within this $z$-band limited sample, we applied a further photometric selection, $K_s \leq 22.2$, to ensure a high fraction of reliable stellar mass ($M_*$) estimates to (see \S \ref{sec:masses}), resulting in a morphologically suitable sample of 1388 objects.  Although spectroscopic redshifts are available for many of our targets from the Team Keck Redshift Survey program \citep{wirth2004} in GOODS-N and from the VIMOS VLT Deep Survey \citep{lefevr2004} in GOODS-S, we did not exclude targets for which only photometric redshifts were available. Selecting within our target redshift range of 0.2$\,<z<\,$1.3, we used photometric redshifts from COMBO17 \citep{wolf2004} and \citet{bundy2005}. As discussed later (\S \ref{compareprevious}), several galaxies within our sample can be found in earlier kinematic surveys of  \citet{flores2006} and \citet{weiner2006b}.

\subsection{Spectroscopic data} \label{sec:spec} \label{spec3}

Over a number of seasons we collected spectroscopic data for this sample with the DEIMOS (DEep Imaging Multi-Object Spectrograph; \citet{faber2003}) instrument on Keck II. In total, we examined 236 galaxies drawn from the target list discussed in \S \ref{sec:sample} (17\% of total sample), simply chosen to maximise number of objects with best position angle alignment with that of the mask. Of the 236 spectra, 129 show line emission extending past what we will term the ``seeing-dispersion beam"\footnote{The seeing-dispersion beam signifies a 2D Gaussian representing the combination of the effect of seeing along the spatial axis and the emission line velocity dispersion along the wavelength axis.}, 59 have only very compact emission lines that sample the central region, and 48 are in a category we will refer to as ``passive"  (meaning galaxies which show no significant emission lines across the 2D spectrum, although weak lines may be recovered in the integrated spectrum) (Fig.~\ref{fig:pasmass_hist}, Table \ref{table_sample}). 

The bulk of our analysis is thus based on the 129 galaxies with extended line emission. While this subsample makes up only 55\% of our initially targeted sample, there is no statistical difference in its redshift or apparent magnitude distribution from the compact and passive subsets (Fig.~\ref{fig:pasmass_hist}). The only obvious differences amongst the 3 subsamples concerns their disk sizes and stellar masses. Disks with extended line emission have larger scale radii than those of the compact and passive subsamples, and passive galaxies are largely drawn from the upper end of the total stellar mass distribution (see Table~\ref{table_sample}). These differences are not surprising and do not lead us to suspect that our working sample of 129 disk galaxies is significantly biased in its range of physical properties compared to the original parent selection. We discuss the properties of our compact emission line sources in a later paper in this series.

The DEIMOS observations were undertaken over a series of runs from March 2004 through April 2008. Slit masks were designed with position angles (PAs) within $\pm$30$^\circ$ of the measured PA in order to minimize tilt angle corrections in the reduction process. We used the 1200 l mm$^{-1}$ grating blazed at 7500 \AA~with 1\arcsec slits (with exception of 7 galaxies observed with the 600 l mm$^{-1}$ grating blazed at 7500 \AA). In this configuration, we achieved a spectral resolution of 1.7 \AA\ corresponding to a velocity accuracy of 30 km sec$^{-1}$. All DEIMOS data were reduced using the automated \textsc{spec2d} pipeline\footnote{http://astro.berkeley.edu/~cooper/deep/spec2d/} developed by the DEEP2 survey.  The \textsc{spec1d} package\footnote{Based on http://spectro.princeton.edu/idlspec2d\_install.html} \citep{davis2003} was used to extract 1D spectra from the rectified 2D spectra produced by \textsc{spec2d}. The combination of 1D and 2D spectra were analyzed using the \textsc{zspec} software, also developed by DEEP2 (Faber et al. in prep, \citet{coil2004}), which fits a linear combination of galaxy, emission line, and stellar template spectra to each spectrum and allows the user to select the best-fitting template, thus determining the spectroscopic redshift. This redshift was used as the initial guess for the systematic velocity in our rotation curve models discussed below. 

Because we are interested in the 2D segments of specific emission lines in the spectra, particular care was taken to separate the reductions of a given mask observed at different hour angles (and therefore different parallactic angles), taken on different runs, or observed under varying conditions.  A scheme (discussed below) was developed to coadd 2D spectra of the same target from multiple observation sets prior to further analysis.  Because the spatial position of an emission line can vary from one observation to the next as a function of wavelength (by $\sim$0\farcs2), we choose to extract $\approx$100 \AA~wavelength ``cutouts'' around key emission lines of interest (the [OII] 3727 \AA\ doublet, H$\beta$, the [OIII] 4959, 5007 \AA\ doublet and H$\alpha$) for our rotation curve study, based on the redshift determined by the \textsc{zspec} analysis. 

Given a set of cutouts for the observation sets of a particular galaxy and emission line, we constructed optimally-weighted coadditions, with weights based on the signal-to-noise (S/N) and seeing full-width-half-max (FWHM) measured from alignment stars on the corresponding slit masks. Typically 5-6 alignment stars were included on each mask. Relatively sky-free windows (with $\Delta \lambda \lesssim 500$ \AA) were selected on both the blue and red sides of each alignment star spectrum.  The stellar flux in these windows was weighted by the inverse variance (as output from the reduction pipeline) and collapsed along the wavelength direction to obtain a stellar profile fitted with a Gaussian.  The width and peak were used to estimate the S/N and FWHM for the blue and red components of each alignment star.  The average FWHM across a mask and the average star-by-star ratio of S/N values provide the seeing FWHM and {\em relative} S/N for that observation set.  The typical spread in FWHM among stars on a given mask is 0.03 arcsec.  The weight of each observation set was then given by $w = s/ f^2$, where $s$ is S/N and $f$ is the FWHM, and these were normalised by the weights of all the coadded masks.

All emission line cutouts were inspected by eye and occasionally rejected if the region extended beyond the detector area or if there was an artifact in the data that could interfere with the line of interest.  To perform the coaddition, each 2D cutout was first rectified to a regular grid in wavelength and spatial position using the 2D wavelength solution output by the \textsc{spec2d} pipeline.  We located the peak of the continuum along the spatial axis by collapsing the 2D cutout in the wavelength direction, initially masking out the emission feature.  The collapsed profile was fit by a Gaussian with the resulting peak taken as the galaxy center.  In rare cases, the continuum was so weak that a position could not be determined without including flux from the emission line itself.  The final centering of each cutout was verified by eye, and the cutouts were coadded with appropriate weighting after alignment in both wavelength and central continuum position.

The seeing varied from 0\farcs6 to 1\farcs2 over the various observing runs, so whenever seeing measurements are needed in the dynamical analysis (\S \ref{sec:analysis}), we adopt the average value measured from the alignment stars in the final coadded observations. This measurement is preferred to that based on a photometric image, since it not only refers to data integrated over the entire exposure time of the spectra, but we also account for systematics in the observing and co-addition process (which use different weights for different exposures).

\begin{table}
\caption{Disk sample: \textsc{Note--} See Fig.\ref{fig:pasmass_hist} for histogram and best gaussian fits of stellar mass and scale radii distribution}
\label{table_sample}
  \begin{tabular}{@{}llllll@{}}
\hline
\hline
Line profile & N & $\langle{\log M_*}\rangle$\footnote{best-fit gaussian centroid of log stellar mass} & $\sigma_{M_*}$ \footnote{\emph{FWHM} of best-fit Gaussian of log stellar mass} & $\langle{r_s}\rangle$\footnote{best-fit gaussian centroid of scale radii in kpc} & $\sigma_{r_s}$\footnote{\emph{FWHM} of best-fit Gaussian of scale radii in kpc} \\
\hline
Extended\footnote{possible to fit rotation curves}&129&10.11$\pm$0.05&0.60 &2.68$\pm$0.09&1.02\\
Compact &59&10.21$\pm$0.10&0.73&2.09$\pm$0.13&1.01 \\
Passive &48&10.69$\pm$0.09&0.62&2.24$\pm$0.16&1.08 \\
\hline
\textsc{total} & 236 & & & & \\
\hline
\end{tabular}
\end{table}


\subsection{Photometric data and bulge-disk decomposition} \label{sec:phot}

By selecting our sample within the GOODS North and South fields, we ensure excellent quality multi-color data for all our galaxies from the Hubble Space Telescope Advanced Camera for Surveys (HST ACS) \citep{giaval2004}. This provides valuable structural information that can be used for translating the observed rotation curves into physically-based properties.  To the imaging in four bands from HST ($B_{435}$, $V_{606}$, $i_{775}$, and $z_{850}$ -bands), we add ground-based K-band data in order to secure stellar mass estimates based on SED fitting (see \S \ref{sec:masses}). 

A key requirement for our analysis is the inclination, PA and
effective radius of each galaxy. We also need to separate, where
possible, the disk light from bulge contamination. We derive these
quantities from the HST imaging using the \textsc{galfit3}
\citep{peng2010} least squares elliptical-fitting method. For each galaxy we extracted a
9\farcs03 x 9\farcs03 (301 x 301 pixels) postage stamp centered on the
object. Neighbors were individually masked out to eliminate confusion.
We first fit an exponential disk component plus a de Vaucouleurs'
bulge profile to every galaxy. Those galaxies which yielded unphysical
solutions were re-fit with a single S\'{e}rsic profile component, where
the S\'{e}rsic index ($n$) was allowed to vary. Such cases generally
represent disk galaxies which are bulgeless and/or more clumpy and
irregular than regular well-formed spirals. $\sim$60\% of our galaxies
were fit using a 1-component $n$-varying S\'{e}rsic profile fit, and
$\sim$40\% were adequately fit with a two-component bulge and disk
solution. This mixture gives some indication of the morphological
distribution of our sample indicating that less than half are
well-formed spirals (\S \ref{sec:sample}). Disk sizes, inclinations
and PAs were taken from best-fit disk component if more than one
component can be fit. To track possible biases in the disk-bulge
decomposition we will later flag those galaxies for which significant
bulge components were measured.

We ran \textsc{galfit} using HST data in all four bands ($B_{F435W}$, $V_{F606W}$, $i_{F775W}$, and $z_{F850LP}$). The disk scale radii are consistent among the bands indicating no significant redshift-dependent bias (or {\it morphological k-correction}) within the sample. In order to maximize the signal-to-noise, hereafter we used the \textsc{galfit} results from the $z_{F850LP}$ band. To achieve convergence on the \textsc{galfit} parameters and to assess any systematic uncertainty in the fitting technique, we ran a Monte Carlo analysis (N=1000) where we varied the initial guess of each parameter (magnitude, effective radius, $b/a$ for inclination, position angle, and sky) from one adopted by the GOODS \textsc{SExtractor} results (\citet{giaval2004}). We found the parameter output distributions were much narrower than the input distributions thereby demonstrating convergence.  Final parameter uncertainties from the Monte Carlo distributions are better than 5\% on average, and we add these uncertainties in quadrature to the observational error.
 
\subsection{Stellar masses} \label{sec:masses}

Reliable stellar masses are an essential component of constructing a baryonic Tully-Fisher relation. We take our stellar mass estimates from work initially presented in \citet{bundy2005}, followed by the analysis presented in \citet{bundy2009}. Further details can be found in those papers.

Briefly, stellar masses are derived using a matched catalog of multi-band ACS and ground-based $K_S$ photometry. The essential near-infrared data was taken with the MOIRCS imager on the Subaru telescope for GOODS-N \citep{bundy2009} and the ISAAC instrument on the ESO VLT for GOODS-S \citep{retzla2010}.  The final matched catalog is substantially complete to a limiting magnitude of $K_{AB}$=23.8, deeper than our spectroscopic limit. 

A Bayesian code fits the spectral energy distribution (SED) derived from 2 arcsec diameter ACS and $K_s$ photometry adopting the best spectroscopic redshift and this SED is compared to a grid of \citet{bruzua2003} models that span a range of metallicities, star formation histories, ages and dust content. The stellar mass is estimated by multiplying the derived K-band mass/light ratio by the observed K-band luminosity derived from the \verb!MAG_AUTO! total Kron magnitude determined by SExtractor.  We assume a \citet{chabri2003} initial mass function (IMF). The probability for each fit is marginalized over the grid of models giving a stellar mass posterior distribution function, the median of which is the catalogued value. At the magnitudes probed in this survey, the uncertainties inferred from the width of these posterior functions is less than 0.1 dex. Including systematic errors (see \citet{bundy2005} for a full discussion), we determine that the stellar masses are reliable to better than 0.2 dex, modulo uncertainties arising from the IMF normalization.

In order to construct a self-consistent TF relation, we apply an aperture correction to the total stellar mass estimates (i.e., for a given fiducial radius, an aperture stellar mass at that radius compared to the velocity measured at that fiducial radius). We extract equivalent Kron radius aperture fluxes on the $z_{F850LP}$ band data to get the flux equivalent to that used in the total stellar mass estimates. We then take a scaled-down aperture flux within the given fiducial radius, and compare this to the Kron radii flux thus deriving a total-to-enclosed flux ratio. Assuming the $z_{F850LP}$-band and $K$-band are roughly equivalent stellar tracers, we can use this ratio to estimate the enclosed stellar mass. This approach maximizes the utility of the HST $z_{F850LP}$ images which have much better resolution than our ground-based $K$-band data, thereby giving us the resolved mass distribution throughout the disk to match the detail seen in our rotation curves. 

\section{Dynamical Analysis} \label{sec:analysis}

We now turn to the questions of how to extract reliable rotation curves from our 2-D spectroscopic data and how to interpret those curves in terms of a fiducial velocity measurement that can be used in the various Tully-Fisher relations we will example. To fully exploit our extended integrations, we re-examine the rotation curve model and consider carefully how to define a self-consistent fiducial velocity that can be robustly measured within our data. We quantify improvements in our data by comparing our velocities (extracted from 6-8 hours of integrated exposure time) with those deduced from $\sim$1 hour subsets, equivalent to exposures made in previous studies \citep{vogt1997,consel2005,weiner2006a,kassin2007}. We also examine the remaining uncertainties given what has been learned from the first studies with IFU spectrographs. In what follows, our analysis is based on the 129 galaxies for which extended emission is observed (\S \ref{sec:spec}).

\subsection{Rotation curve model} \label{sec:model}

\begin{figure}
\center
   \includegraphics[width=3.4in]{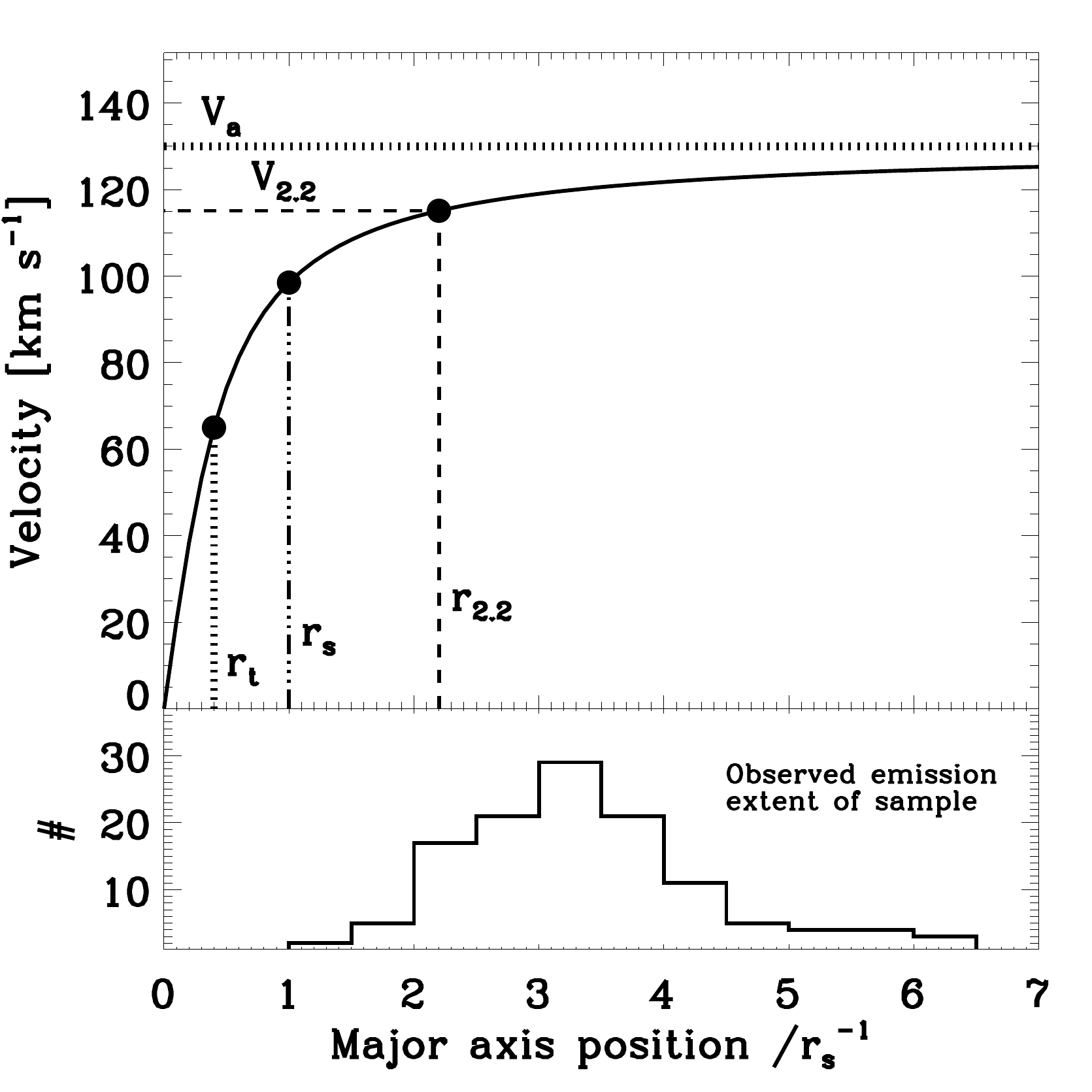}
   \caption{A functional arctan rotation curve with various characteristic radii (discussed in the text) shown as dimensionless factors of the disk scale length ($r_s^{-1}$). The example shown has a turnover radius, $r_{t}\sim0.4\,r_s$, and an asymptotic velocity, $V_a\sim130\,km s^{-1}$, typical of many galaxies in our sample. The region where the rotation curve turns over is extensive and the so-called turnover radius, $r_t$, does not necessarily indicate the most appropriate center of this region for the arctan function. The histogram below shows the extent to which we can reliably trace emission lines in the spectra of our sample. The frequently used $V_{max}$ (maximum measured velocity) is not equivalent to the asymptotic value, $V_a$, a mathematical extrapolation and typically not reached in the observed rotation curve. Estimates of $V_a$ can depend critically on how well-constrained are the central region and turn over of the rotation curve. As $\sim$90\% of our galaxies are traced to 2.2$r_s$ (and all to at least 1.0$r_s$), in our TF relations we opt to use $V_{2.2}$, the velocity at $r_{2.2}$, which minimizes uncertainties arising from extrapolation (see \S \ref{sec:model}). }
   \label{fig:mod}
\end{figure}

\begin{figure*}
   \centering
   \epsscale{1.1}
   \plottwo{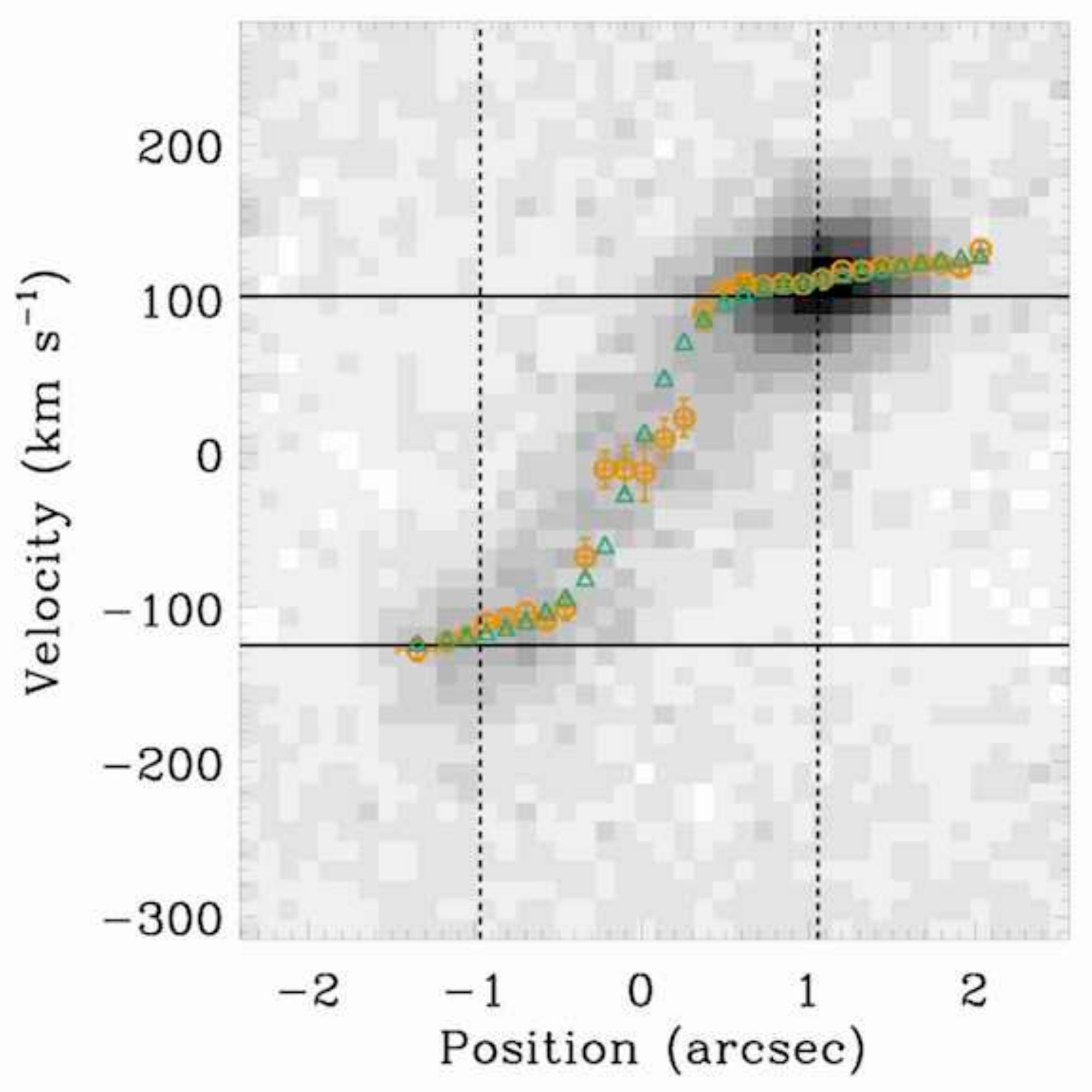}{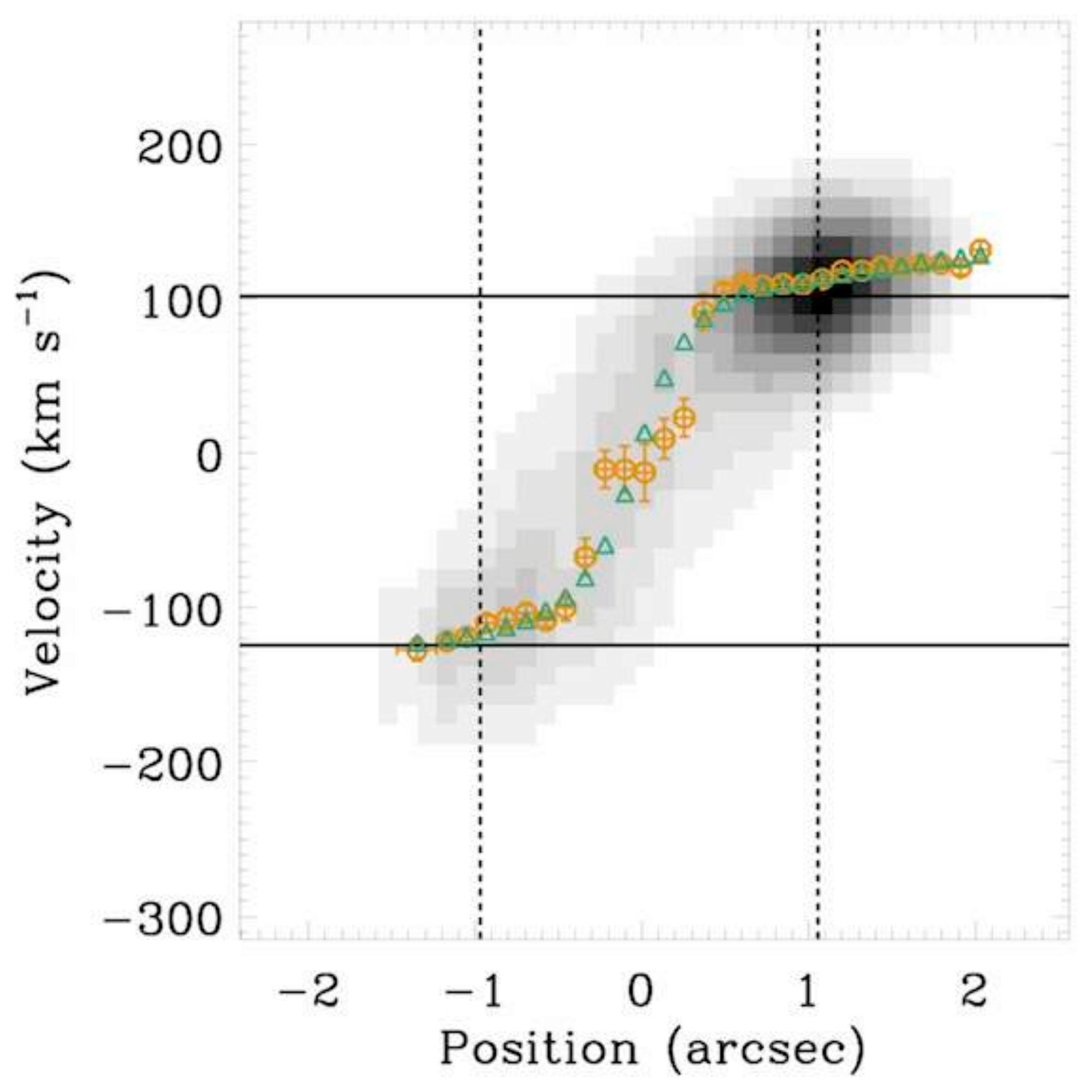}
   \caption{An example of our fitting procedure for a galaxy with asymmetric emission. The left panel shows the
     observed 2-D spectral data in the region of $H\alpha$, and the right panel is the equivalent
     section of the constructed model spectrum. Both show a grey scale representing flux as a
     function of velocity with respect to position. Orange circles show the trace centroids of the
     emission line (bars along the positional dimension indicate bin size, not error), and the green
     triangles show the equivalent centroids for the constructed model. The $r_{2.2}$ radius is
     marked by vertical dotted lines and the extracted $V_{2.2}$ velocity is marked by horizontal
     solid lines. These lines of fiducial measurements do not cross directly through the trace
     because they are extracted from the best-fit model's intrinsic arctan function, prior to distortion 
     by seeing and dispersion. See \S \ref{sec:fit} for more details.}
   \label{fig:tracemod}
\end{figure*}

Optimally fitting rotation curves presents a variety of challenges that become more difficult as redshift increases. Foremost, we seek a functional form which represents the bulk of the observed emission line shapes and has some physical basis. Secondly, we must aim to characterize this functional form with a fiducial velocity that is reliably detected across the sample, preferably without extrapolation to radii where there is no data. Given our extended integrations, we have considered carefully the optimum selection of this characteristic velocity. The challenges can be appreciated by considering Figure \ref{fig:mod} where we show various characteristic velocities in the context of the frequently-used arctan model of a rotation curve \citep{courte1997} as well as the extent to which we can trace emission lines for our sample. 

 Several functional forms have been discussed in the literature, such as the ``multi-parameter function'' in \citet{courte1997} and the ``universal rotation curve'' of \citet{persic1996}. The simplest model flexible enough to fit most rotation curves is the empirically-motivated arctan function (see Fig. \ref{fig:mod}), which we adopt here, viz:
\begin{equation}
V = V_{0} + \frac{2}{\pi} V_{a} \arctan(\frac{r - r_0}{r_t}),
\label{eq:court}
\end{equation}
where $V_0$ is the central or systematic velocity, $r_0$ is the dynamic centre, $V_{a}$ is the asymptotic velocity, and $r_t$ is the turnover radius, which is a transitional point between the rising and flattening part of the rotation curve \citep{courte1997, willick1999}.  The arctan model does not account for a sharp peak that is found in some local, bulge-dominated rotation curves around the turnover radius, but we typically do not observe this feature in our sample. 

Past studies employed as TF velocities the maximum measured velocity along the disk, $V_{max}$, or the asymptotic velocity from Eq.~\ref{eq:court}, $V_a$ \citep{vogt1996,vogt1997,weiner2006a,kassin2007}. The disadvantage with $V_{max}$ is that it is not measured at a consistent location in the variety of disks observed. In terms of the disk scale radius, we can see a range of a factor $\simeq$5 or so in the associated radius.

Some studies \citep[i.e., ][]{flores2006,weiner2006b} advocate the use of $V_{circ}$, the circular radial velocity, driven by either the disk or the halo, but in nearly all cases this is assumed to be the $V_a$ of the widely adopted arctan model. While the arctan function closely matches the observed extent of rotation curves, the mathematically-extrapolated asymptote is unwarranted as it is not observed in typical datasets, even for our extended integrations. Small offsets in the extrapolated velocity curve will lead to large changes towards the asymptotic limit, especially when emission is not detected past the flattened part of the rotation curve, and emission line dispersion and seeing distort the terminal emission. Under these circumstances, the majority of the `$V_{circ}$' velocities remain as extrapolations. 

\begin{figure*}
   \centering
     \includegraphics[width=6in]{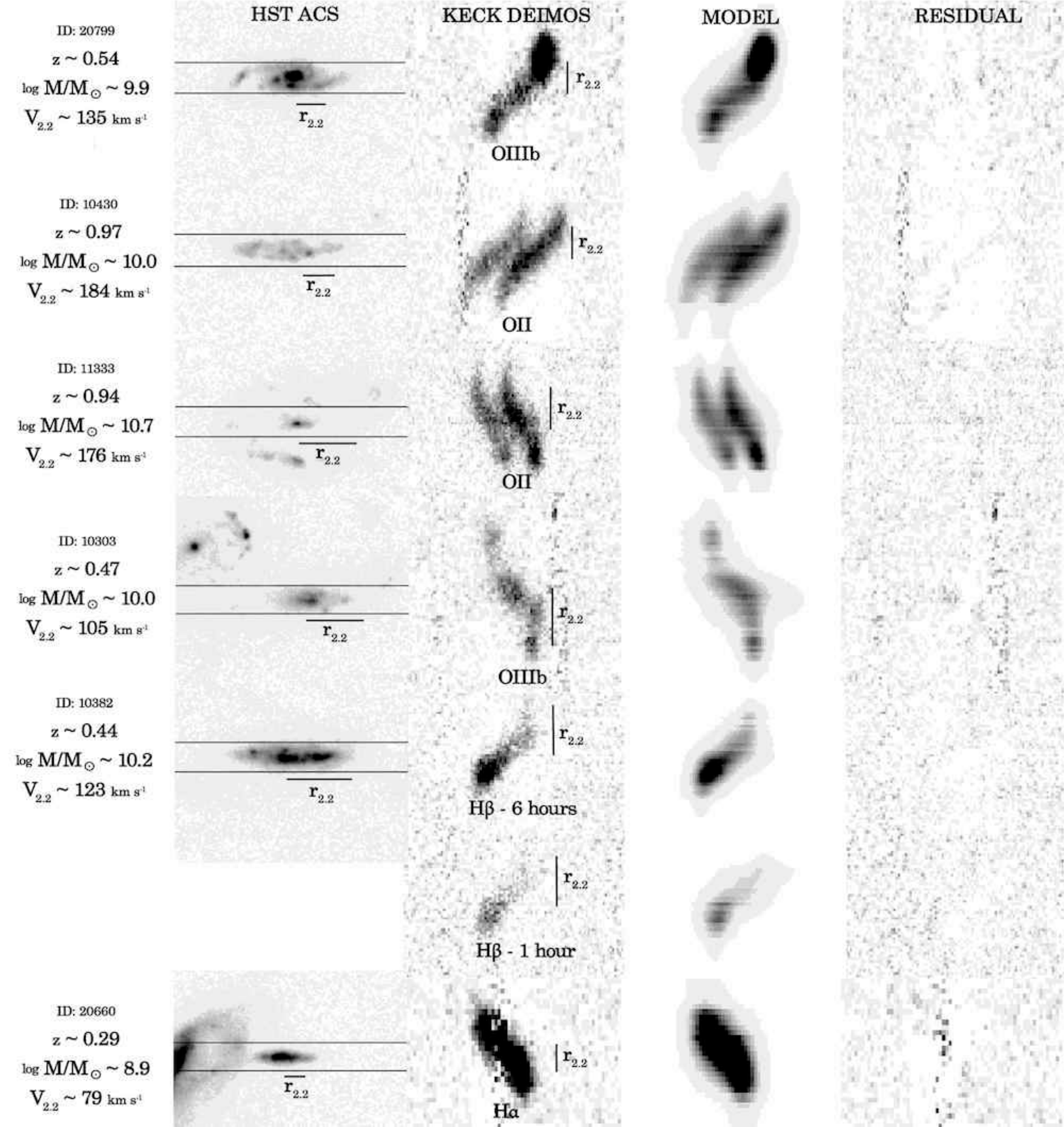} 
   \caption{Examples of data from the survey. For each source we show an HST ACS composite ($B,V,i,z$) photometry postage stamp rotated for convenience so the overlaid 1 arcsec slit level is horizontal, the Keck DEIMOS 2-D emission line cutout, the corresponding best model and residual of model minus data. The extent of $r_2.2$, where $V_{2.2}$ is measured is also overlaid. The triptych second to the bottom shows the same emission line as the one above but with only 1 hour spectroscopic integration.}
   \label{fig:ex}
\end{figure*}

We thus seek a physically motivated fiducial radius to which emission can be detected across the variety of disks seen over our wide redshift range.  We have adopted the modeled velocity at 2.2 times the exponential disk scale length, $r_{2.2}$, which we will call $V_{2.2}$. This has a good physical basis as it corresponds to the location of peak rotational amplitude for a pure exponential disk \citep{freema1970, binney1987, courix1997}. Although few disks are likely to be perfectly exponential, the $r_{2.2}$ approximation as the point at which rotation curves flatten is visually confirmed in most of our galaxies. TF velocities based upon $V_{2.2}$ result in the smallest internal scatter and provide the best match to radio (21 cm) line widths for local galaxies \citep{courte1997}. This fiducial velocity has been adopted by \citet{dutton2010b} in determining the kinematic connection between late-type galaxies and dark matter halos. As shown in Figure \ref{fig:mod} we trace the velocity field with at least one optical emission line to $r_{2.2}$ for  $\sim$90\% of our sample (Fig. \ref{fig:mod}).

\subsection{Rotation curve fitting procedure}\label{sec:fit}

\begin{figure*}
   \centering
     \includegraphics[width=6in]{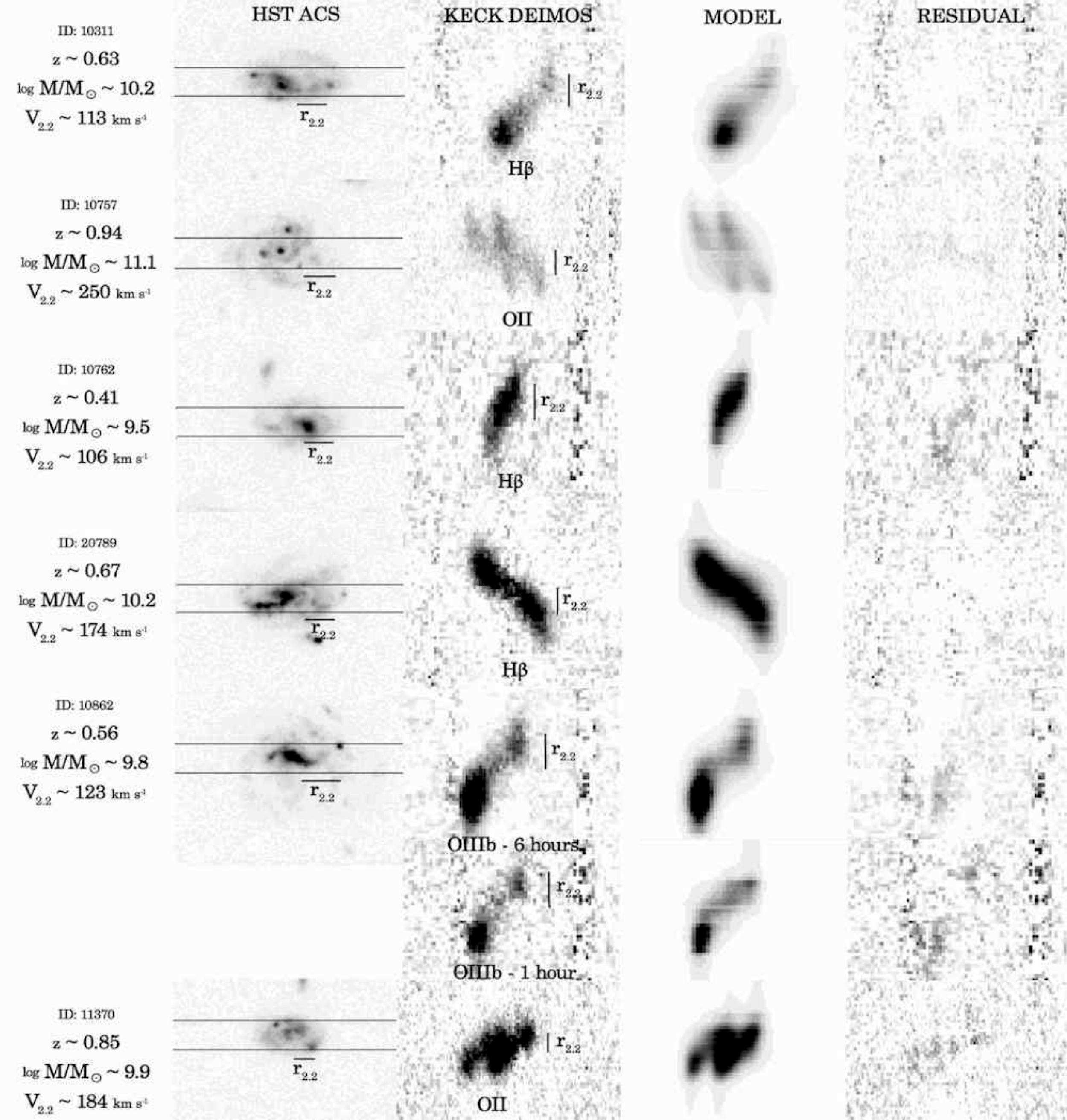} 
   \caption{As Figure \ref{fig:ex} but for more disturbed, visually asymmetric disks, some of which appear to be undergoing minor mergers. Although the morphologies are more irregular, we usually succeed in fitting an appropriate arctan-based model.}
   \label{fig:ex-disturb}
\end{figure*}

We now discuss how we fit the emission line data to obtain an accurate
velocity measurement at $r_{2.2}$ for each galaxy disk using the
arctan model. In the next section we discuss various inclination
corrections that must be made following the fitting procedure. A
simple outline of our fitting code is as follows:

\begin{itemize}
\item Trace the observed 2D emission line in wavelength as a function
  of spatial position,
\item Construct a model 2D spectrum with an arctan-shaped
  emission line profile, implementing features
  measured from step 1 (e.g., position-dependent emission brightness
  profile, dispersion, and seeing),
\item Trace the model spectrum, varying the arctan model parameters
  $V_{a}$ and $r_{t}$ (as well as the dynamical center and a
  seeing-dispersion beam correction factor) until the model trace
  optimally fits that of the data,
\item Compute $V_{2.2}$ at $r_{2.2}$ from the best-fit arctan models.
\end{itemize}

In detail, our procedure is as follows:

\emph{Step 1: }We begin by subtracting the galaxy continuum on the
rectified 2D spectral frame, leaving only the emission-line profile.
This is performed by linearly interpolating in wavelength across the
emission line region, treating each spatial row independently.

\emph{Step 2: }We next trace each emission line as a function of
spatial position. This procedure returns the central wavelength, line
width, and peak flux in each spatial bin. We fit two half-Gaussians 
to the line profile in the wavelength direction, and use an adaptive binning procedure in
the spatial direction to ensure a consistent signal-to-noise (S/N).
The blurring effect of seeing in the spatial direction, and of
dispersion in the wavelength direction, can affect the
emission line in a particular spatial bin by mixing flux from neighboring bins. We use
two half-Gaussians with differing sigmas joined at the same peak to account for these effects, 
ensuring that the position of the peak flux is always traced.

The trace terminates when the emission is no longer detected above the
local noise level. All traces are visually inspected to ensure that
spurious reduction artefacts do not interfere with the fit. The
procedure produces an array of wavelength values as a function of
spatial position (see Fig.~\ref{fig:tracemod}), as well as the
emission peak flux. 

We also experimented with different forms for the emission line
brightness profile along the slit, for example the collapsed light
through a synthetic slit estimated from the HST image, or an
exponential profile. However, traces using these profiles resulted in
larger residuals.

\emph{Step 3:} We next construct a model 2D spectrum from the rotation
curve of eqn.~(\ref{eq:court}), which requires $V_{a}$ and
$r_{t}$ parameters for the arctan functional form. We guess initial
values to approximately match the flat portion of the outer trace of 
the observed data. We lay this model rotation curve on a 2D grid, adjusting
the normalisation (brightness) and line width in each spatial bin to
match that of the data (i.e., using the parameters from the
two half-Gaussians in each bin). We finally convolve the resulting
rotation curve with the local seeing (see \S \ref{spec3} for details),
and the resulting peak model brightness is re-normalized, bin by bin,
to that observed.  Ideally we would start with the unblurred emission-brightness spatial
profile, but there is a degeneracy in blurring by seeing and dispersion, 
which is particularly troublesome for galaxies with irregular emission
line brightness profiles.

Because we have measured the position-dependent dispersion from the
trace (which implicitly contains the effect of the seeing), as well as a
separate measurement of the seeing from the alignment stars, we can 
attempt to break this degeneracy by fitting for a multiplicative factor. 
This correction factor is used to multiply the dispersion implemented in the model,
and it is always less than or equal to 1. This allows us to approach the intrinsic, 
deconvolved rotation curve. This method is more successful than 
alternative approaches we attempted using both synthetic and observed data, 
for example adopting the collapsed light profile from the most appropriate 
broad band HST image as observed through the slit, or simple functional 
forms of emission brightness 
profiles (constant, linear, exponential). Indeed, we find many emission lines that
are brighter at the disk edge than towards the center, as well as
asymmetric emission distributions which do not match the broad-band
flux distributions that enter the slit. 


\emph{Step 4:} To select the best model, we use a robust non-linear
least squares fitting algorithm based on \textsc{mpfit}
\citep{markwardt2009}. The arctan input parameters, $V_{a}$ and $r_t$
are varied as well as the position of the dynamical center and the
seeing-dispersion beam correction factor.  Chi-squares are calculated
between the model trace and the data trace, rather than on the entire
2D frame. Not only is this method much faster, but we find it results
in smaller residuals, simply because we are focusing the fit on the
information that is most important to the shape of the rotation curve.
We propagate the error from the measured input parameters (emission
profile, seeing, and dispersion) by using a Monte Carlo approach,
simultaneously altering input parameters from random Gaussian
distributions with widths matching the error in the observed
parameters for each galaxy. After 100 iterations, we add the
uncertainty from the resulting distribution of output velocities to
the formal fitting errors.

\vspace{0.1cm}

The velocity at $r_{2.2}$ can then be calculated using the best-fit
arctan model and the radius measured from the $z_{F850LP}$ band HST
data (see \S \ref{sec:phot}). All the observed emission lines in a
given spectrum are treated independently in the above steps. These
independent measurements of $V_{2.2}$ are found to be consistent in
$>90\%$ of our galaxies. We combine the emission line fits for the
same galaxy into a final weighted average of $V_{2.2}$ and include the
error in $r_{2.2}$ in the final uncertainty.

Examples of our spectral data and respective best-fit model rotation curves can be
found in Fig.~\ref{fig:ex} and Fig.~\ref{fig:ex-disturb}. Note that these 
models represent the minimum chi-square best-fit on the 1-D trace of the 2-D 
models as described above, not a direct 2-D chi-square best-fit. Fig.~\ref{fig:ex} 
shows six examples of disks which appear mostly morphologically-regular in their HST multi-filter
photometry, whereas Fig.~\ref{fig:ex-disturb} displays six galaxies which are
likely more disturbed or irregular based on their HST images. In both figure sets,
we observe fairly regular arctan-shaped kinematics, although places
of high dispersion and brightness in the emission line tend to coincide with regions
of the galaxy in the HST image that appear to be internally or externally disturbed. For the
5th galaxy in both figures, we show the best-fit model rotation curve after only 1 
hour of integration time for a comparison to the full 6-8 hours of integration shown in 
the panel just above.

\subsection{Inclination and PA offset corrections}
\label{sec:incl-pa-offs}

We now correct our $V_{2.2}$
measurements for the effects of disk inclination and any misalignment
between the position angle (PA) of the DEIMOS slit and the major axis
of the galaxy as determined from GALFIT.

Adopting the convention $i=0^\circ$ for face-on and $i=90^\circ$ for
edge-on disks, the inclination correction is:
\begin{eqnarray}
V_{corr} &=& \frac{V_{obs}}{(\sin{i})}, \\ 
i &=& \cos^{-1} \sqrt{\frac{(b/a)^2 - q_0^2}{1- q_0^2}},
\label{eq:incl}
\end{eqnarray}
where $q_0$$\simeq$0.1-0.2 represents the intrinsic flattening ratio
of an edge-on galaxy \citep{haynes1984,courte1996,tully1998}. Although
the precise value depends on morphology, the uncertainty leads to
changes in the final velocity measurement on the order of 1 km\,s$^{-1}$
\citep{pizagn2005,haynes1984}. We assumed $q_0$=0.19 for all
systems, similar to \citet{pizagn2005}.

For the PA offset, we determine a correction: 
\begin{equation}
V_{corr} = \frac{V_{obs}}{\cos{(\Delta PA )}}.
\label{eq:pa}
\end{equation}
from simple trigonometry accounting only for 
the misaligned slit component of the true major axis. Only 10 galaxies in our
final sample have velocity corrections arising from a PA offset
greater than 10\%, and no PA offset exceeds $45^\circ$.

We apply these two corrections to produce a catalog of corrected
$V_{2.2}$ measurements, where the effects of seeing, velocity
dispersion and the emission line brightness profile are accounted for
in the model. As explained in \S \ref{sec:masses}, to compare like
with like in our TF relations, we match $V_{2.2}$ to an estimate of
the stellar mass within $r_{2.2}$. In order to compare this
\emph{enclosed} relation to a more familiar construction of the TF
relation used in previous studies, we also consider a \emph{total}
relation, which compares the stellar mass estimated from the Kron
radius with the velocity associated with the optical radius $r_D$,
defined for an exponential profile as one enclosing $\sim$83\% of the
light, or 3.2$r_s$. This measurement of $V_{3.2}$ or $V_{D}$ we find
to be comparable to the $V_{circ}$ or $V_{max}$ of previous
intermediate-redshift studies, however our choice of $r_D$ is more
consistent between disks than the $r_{max}$ used in those studies.

\begin{figure*}
   \centering
   \plottwo{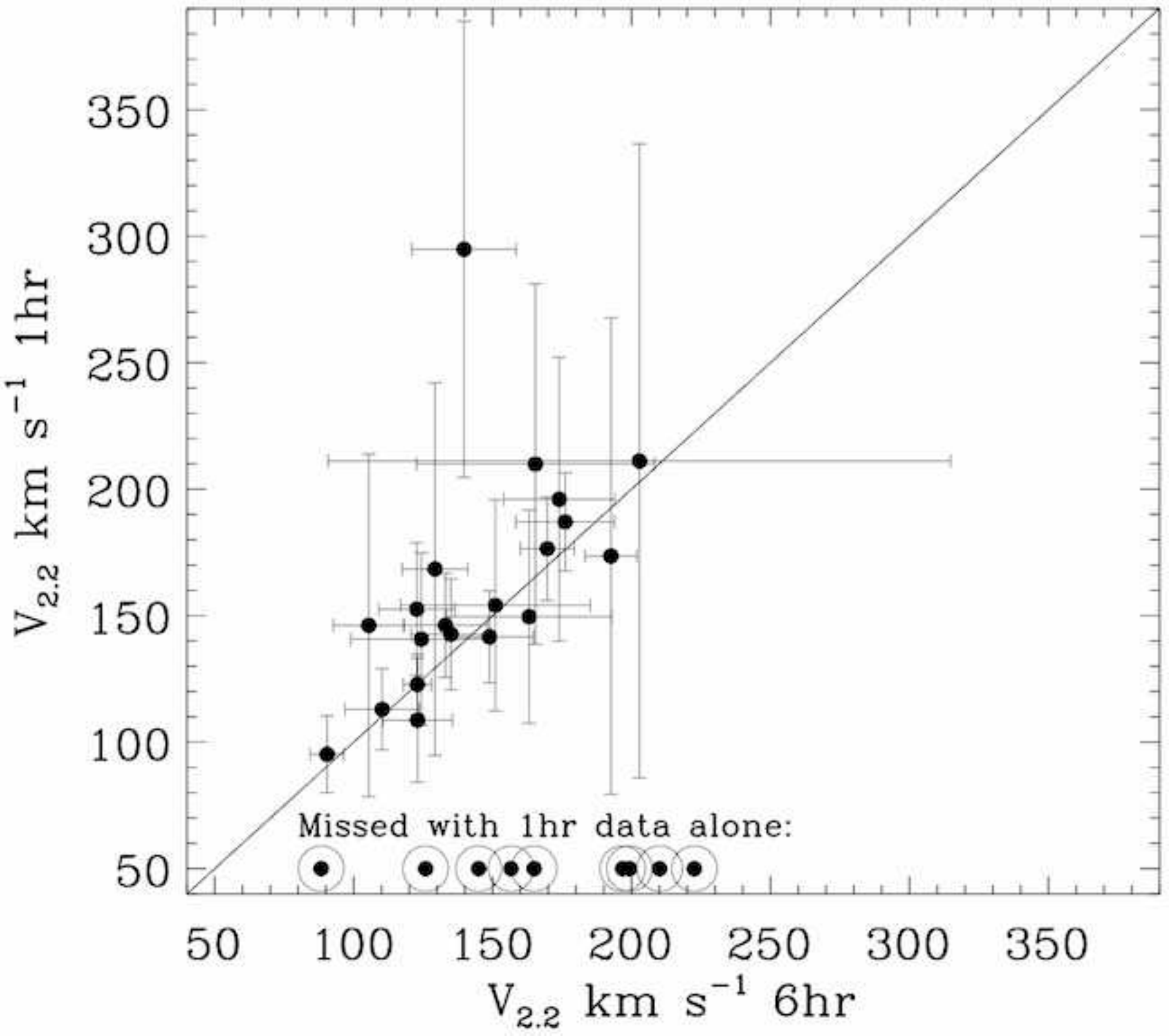}{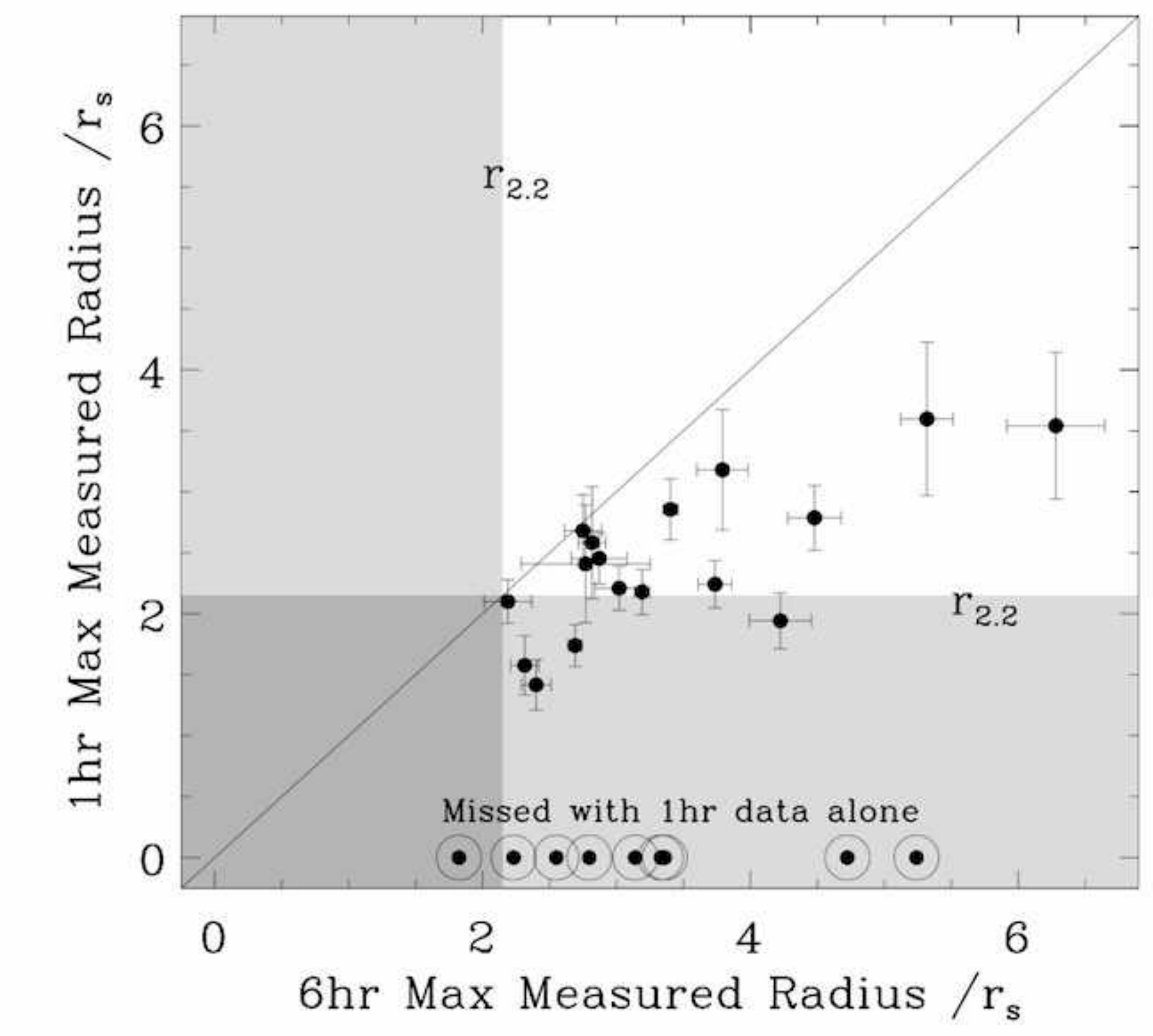}
      \caption{A comparison of the modeled $V_{2.2}$ values and maximum measured emission line extent as derived from our total exposure time (6 hours in these examples) to those derived from a 1 hour exposure extracted as a subset of our data. Left: the agreement between $V_{2.2}$ values is reasonable given the errors, however there is a systematic offset and the errors are significantly larger for the 1 hour subset. Right: almost every galaxy reveals significantly more extended emission in the 6 hr data. The shading marks the $r<r_{2.2}$ region; if the emission does not extend beyond this region, $V_{2.2}$ must be extrapolated, as is the case for several 1hr galaxies. Note also that a third of the rotation curves cannot be adequately traced with only 1 hour of integration (circled points).}
   \label{fig:v_1_vs_6}
\end{figure*}

\subsection{Demonstrating the advantages of extended integrations}\label{onesix}

A major advance in our survey is the use of extended integration times (typically 6-8 hours) on sources with apparent magnitudes ($z_{F850LP}<$ 22.5) that have typically been observed for $\sim$1 hour exposure time {\citep{vogt1997,consel2005,weiner2006a,kassin2007}. In addition to ensuring our emission lines are traced to $r_{2.2}$ (Figure \ref{fig:mod}), this leads to improved signal/noise at all radii. We can demonstrate the effect this has on our derived velocities by using our model fitting code on a subset of our data taken with a 1 hour integration time.  Comparing the results with the 6 hour integrations in Figure~\ref{fig:v_1_vs_6}, we can draw two important conclusions.

Firstly, while most $V_{2.2}$ measurements are consistent given the
error bars, the 1 hour data has significantly larger error bars and
there is a systematic offset whereby the 1 hour analysis produces on
average a 13\% higher measurement of $V_{2.2}$ than the 6 hour data.
This suggests that if errors are not properly accounted for, studies
using a similar modeling procedure but with less integration time may
falsely detect evolutionary signals as a result of over-estimated
velocities due to the decline in signal-to-noise.

Secondly, as expected, the 6 hour data enables us to probe largely beyond $r_{2.2}$, whereas this is not the case for the 1 hour subset. The right hand panel in Fig.~\ref{fig:v_1_vs_6} shows the gains made in detected emission extent in factors of scale radius. We trace, on average, 30\% further along the disk with the 6 hour data than the 1 hour data. In fact, modeling an arctan function fails for one third of the 1 hour sample due to the low signal-to-noise, and missing segments of the emission create a false deviation from the arctan shape. We are confident that we have successfully converged on necessary integration time in our 6 hour data because emission is traced beyond $r_{2.2}$ in $\sim$90\% of our disks, ensuring we can accurately measure $V_{2.2}$.

\subsection{Comparison to previous work}
\label{compareprevious}

Our rotation curve fitting method was constructed to work on data with
extended exposure times, so it is interesting to compare to velocities
for the same galaxies derived using alternative techniques. The 
greatest overlap can be found using the TKRS/GOODS sample \citep{wirth2004,giaval2004}. 
Figure~\ref{fig:newcomp} shows 35 of our
galaxies in common with in the TKRS/GOODS sample for which there are 2D spectral fits for
$V_{rot}$ (equivalent to $V_{a}$ and does not include a correction for
inclination), and all of our sample overlaps with their 1D linewidth
measurements \citep{weiner2006b,weiner2006a}. We compare the velocities we measure
for these galaxies without the inclination correction applied for a most direct comparison.

\begin{figure}
   \centering
     \includegraphics[width=3.5in]{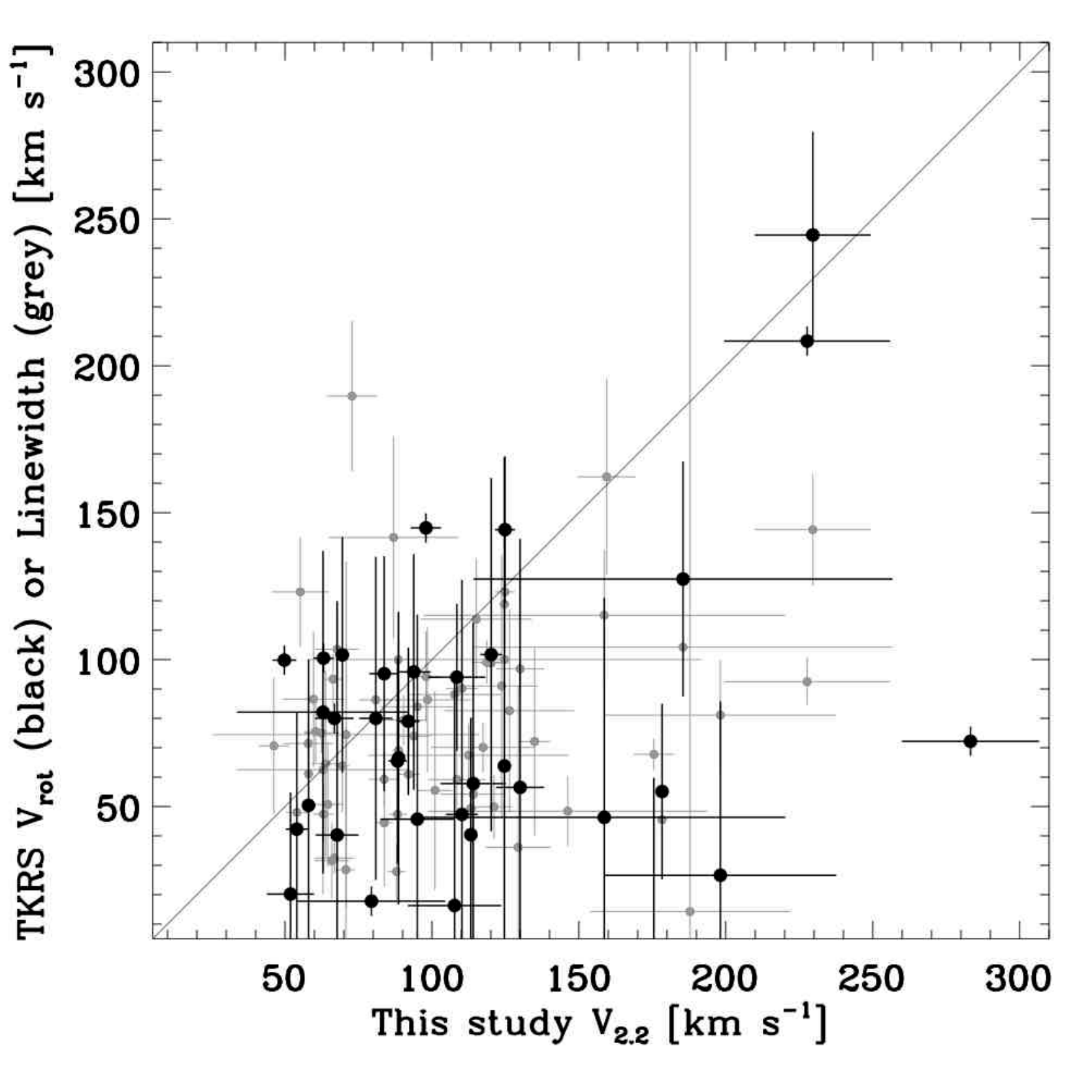} 
     \caption{A comparison of the velocities for 35 galaxies that
       overlap between this study and the TKRS/GOODS study of  \citet{wirth2004} \& \citet{giaval2004}. In
       black are shown the The TKRS/GOODS 2D spectral fits for
       $V_{rot}$ (equivalent to $V_{a}$), and in gray are the TKRS/GOODS 1D linewidth
       measurements. Inclination corrections are not applied to any of the velocity measurements plotted here.}
   \label{fig:newcomp}
\end{figure}

On average, the mean $V_{rot}$ for the TKRS 2D data is 68\% of the
mean $V_{2.2}$ for our study. For the 1d linewidth data, the TKRS
measurements have a mean that is 75\% of the $V_{2.2}$ mean in our
study. Our current data have significantly smaller error-bars as would
be expected given the longer integration times: the median fractional
error bar is 8.6\% for this study, and 59.1\% for TKRS.

The three galaxies in common with the \citet{flores2006, puech2008}
studies have consistent velocity measurements within the error bars
when comparing only the equivalent slit area from our study to their
modeled IFU data. One galaxy has an inconsistent TF velocity
measurement since a higher $V_{max}$ is found on either side of the
slit area on the full modeled IFU velocity field. Little can be concluded from a
comparison of three objects, however in terms of sample selection, 
it is encouraging that the three shared objects between our studies 
all belong to the sub-class of \emph{Complex Kinematics} in the \citet{flores2006} 
kinematical classification scheme, consisting of objects with the most 
irregular, peculiar observed kinematics.

\section{Results}\label{sec:reses}

\begin{figure*}[h]
   \centering
     \includegraphics[width=5.3in]{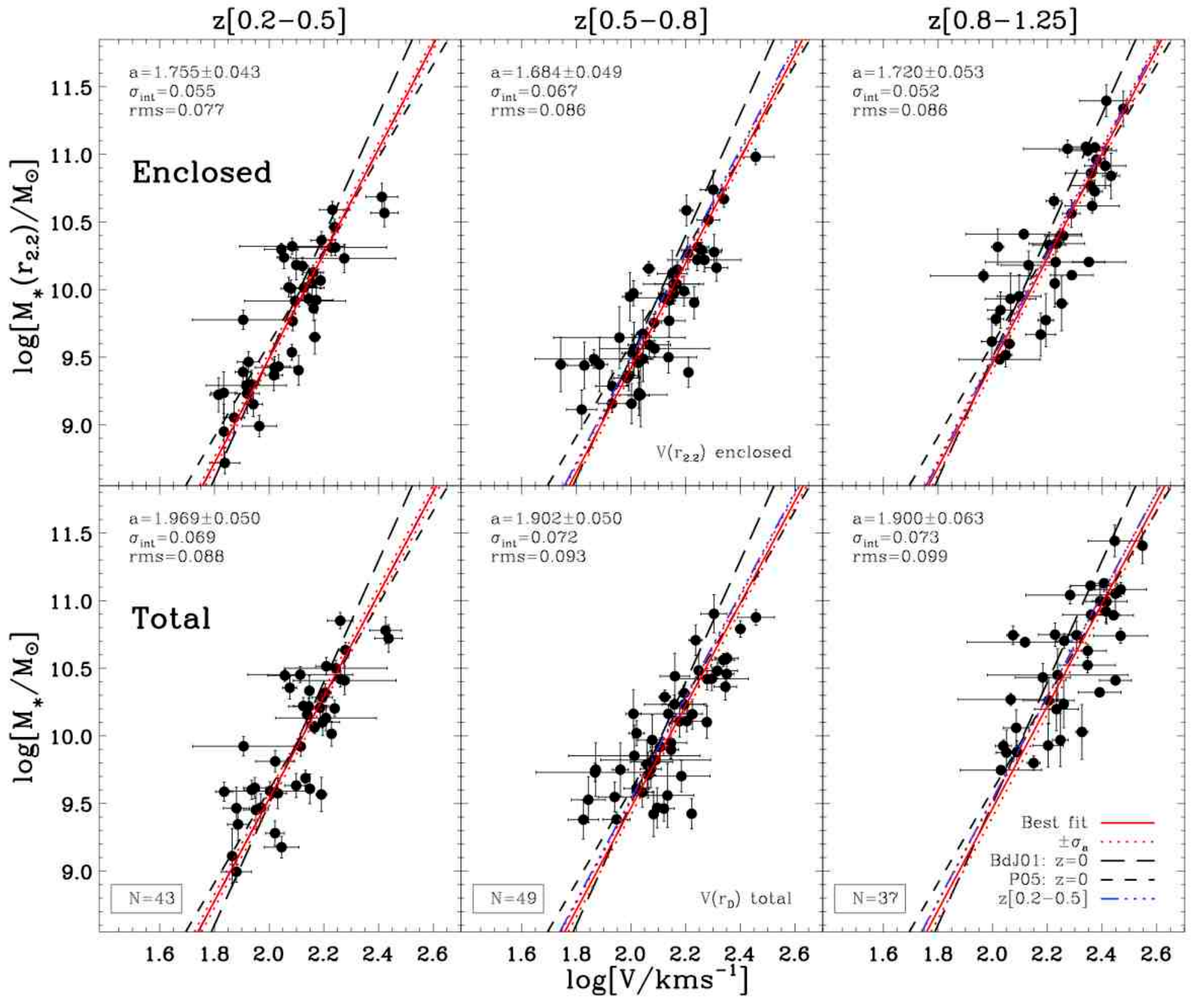} 
      \caption{Redshift-dependent stellar mass Tully-Fisher ($M_{\ast}$-TF) relations using the enclosed (top panels) and total (bottom panels) mass estimates.  Dashed lines refer to the local relations of \citet{bellde2001} and \citet{pizagn2005}. Dot-dashed lines refer to the fit to the lowest redshift bin in the current dataset. To fit fixed slopes between redshift bins, we fit relations using a Monte Carlo distribution of slopes from the best-fit free slope of the entire sample. The resulting mean slope is shown as a solid red line with red dotted lines showing the 1$\sigma$ error in the zero-point (in $M_{\ast}/M_{\odot}$ dex). Using the velocity and enclosed mass at the $r_{2.2}$ aperture reduces both the intrinsic scatter ($\sigma_{\mathrm{int}}$) and the rms of the relationship in each redshift bin.}
   \label{fig:clas_tf}
\end{figure*}

\begin{table*}[h]
 \begin{tiny}
 \caption{Stellar Mass Tully-Fisher Relations}
\begin{tabular}{ccc}
\hline
 \end{tabular}
\begin{tabular}{lllllllllll}
\hline
z range & $\langle z \rangle$ &  N &  $a$\footnote{best-fit y-intercept in $M_{\ast}/M_{\odot}$ dex assuming scatter in $V/km s^{-1}$ dex} & $b$\footnote{slope assuming scatter in $V/km s^{-1}$ dex} & $\sigma_{int,V}$\footnote{internal scatter in $V/km s^{-1}$ dex} & median $\sigma_{V}$\footnote{median velocity error in $V/km s^{-1}$ dex} & $rms_V$\footnote{total scatter in $V/km s^{-1}$ dex} & $\sigma_{int,M}$\footnote{internal scatter in $M_{\ast}/M_{\odot}$ dex}  & median $\sigma_{M}$\footnote{median stellar mass error in $M_{\ast}/M_{\odot}$ dex} & $rms_M$\footnote{total scatter in $M_{\ast}/M_{\odot}$ dex} \\
[0.5ex]
\hline
Enclosed: $M_{\ast}$($r_{2.2}$) vs  V($r_{2.2}$): \\
[0.5ex]
\hline
\textbf{0.2$<$z$\le$1.3} & \textbf{0.64} & \textbf{129} & \textbf{1.718$\pm$0.415} & \textbf{3.869$\pm$0.193} & \textbf{0.058} & \textbf{0.022} & \textbf{0.083} & \textbf{0.224} & \textbf{0.091} & \textbf{0.323} \\
0.2$<$z$\le$0.5 & 0.37 & 43 & 1.755$\pm$0.043 & " fixed & 0.055 & 0.035 & 0.077 & 0.211 & 0.081 & 0.297 \\
0.5$<$z$\le$0.8 & 0.62 & 49 & 1.684$\pm$0.049 & " fixed & 0.067 & 0.045 & 0.086 & 0.257 & 0.118 & 0.334 \\
0.8$<$z$\le$1.3 & 0.96 & 37 & 1.720$\pm$0.053 & " fixed & 0.052 & 0.062 & 0.086 & 0.202 & 0.098 & 0.331 \\
[0.5ex]
\hline
Total:  $M_{\ast}$ vs  V($r_D$) \\
[0.5ex]
\hline
\textbf{0.2$<$z$\le$1.3} & \textbf{0.64} & \textbf{129} & \textbf{1.926$\pm$0.476} & \textbf{3.783$\pm$0.223} & \textbf{0.070} & \textbf{0.023} & \textbf{0.093} & \textbf{0.266} & \textbf{0.087} & \textbf{0.353} \\
0.2$<$z$\le$0.5 & 0.37 & 43 & 1.969$\pm$0.050 & " fixed & 0.069 & 0.037 & 0.088 & 0.262 & 0.083 & 0.332 \\
0.5$<$z$\le$0.8 & 0.62 & 49 & 1.902$\pm$0.050 & " fixed & 0.072 & 0.057 & 0.093 & 0.272 & 0.110 & 0.351 \\
0.8$<$z$\le$1.3 & 0.96 & 37 & 1.900$\pm$0.063 & " fixed & 0.073 & 0.064 & 0.099 & 0.277 & 0.084 & 0.375 \\
[0.5ex]
\hline
\hline
\end{tabular}
\label{table_smtfr}
\end{tiny}
\end{table*}

We now reach the primary aim of our paper: to present the redshift-dependent stellar mass Tully-Fisher ($M_{\ast}$-TF)
relation over 0.2~$<z<$~1.3. We will explore the quantitative improvement, which we have realized through our extended integrations and subsequent rotation curve modeling, with respect to earlier work in a number of ways. Foremost, the scatter around the redshift-dependent relations will provide a good indication of our progress. We will examine relations using masses derived within our chosen fiducial 2.2 scale radius, $r_{2.2}$, the so-called {\it enclosed } relations alongside those for the more traditional {\it total} relations (see \S \ref{sec:masses} and the end of \S \ref{sec:model} for more details). It is also convenient to examine and discuss the $B$-band luminosity-based TF relation from our survey as the literature contains many estimates of this scaling relation and previous studies have claimed evolution, despite large scatter and possible incompleteness biases (\S \ref{sec:bmtfr}). 

\subsection{The redshift-dependent stellar mass TF relation}\label{sec:tfrs}

We begin by considering the case for evolution in the stellar mass ($M_{\ast}$)-TF relation. The results are illustrated in Figure~\ref{fig:clas_tf} and listed in Table~\ref{table_smtfr}. Redshift bins were selected to ensure nearly uniform samples over our total redshift range. Shifting the boundaries of these bins by modest amounts does not change the overall conclusions we present below.

To fit a linear regression to our data we adopt a least-squares approach which incorporates a measurement of the intrinsic scatter $\sigma_{\mathrm{int}}$, which is added in quadrature to the velocity dimension. We fit a zero-pointed line:
\begin{equation}
\log{(M_{\ast})}  = [a + b \log{(V_{2.2})}] - \log{(M_0)},
\label{eq:line}
\end{equation}
where $M_0 = 10^{10} M_{\odot}$, and while we plot the relation in the familiar way with velocity on the x-axis and a y-intercept given in terms of stellar mass, we treat velocity as the dependent variable in the linear regression. Fitting linear regressions with stellar mass as the dependent variable leads to fits which suffer much more from the effects of incompleteness bias \citep{bamfor2006, weiner2006b, kelly2007}. 

To fit the data to Eq. \ref{eq:line}, we adapted a code which takes into account errors in both the ordinate and abscissa. We first fit an unrestricted slope to the entire, un-binned log $V$ - $M_{\ast}$ data set, and then fit the individual redshift-binned relations with a distribution of slopes drawn from a Monte Carlo (N=100) Gaussian distribution of slopes centered on the slope found for the full sample.  We undertake this exercise for both the {\it enclosed} mass (i.e. that associated with our fiducial radius, $r_{2.2}$), and the {\it total} mass. We also experiment with simpler approaches to linear regressions, which result in similar trends but naively smaller errors, so we present here our most robust results with the broadest consideration of the uncertainties.

For a local comparison, we consider the $M_{\ast}$-TF relation derived from $K$-band luminosities published by \citet{bellde2001} and the $M_{\ast}$-TF relation from \citet{pizagn2005}, the latter of which is based on velocities from $r_{2.2}$ as presented in the enclosed relation of this study. 

Before discussing possible evolution, we consider the derived scatter around the relations since this is a valuable indication of our gains in precision. Satisfactorily, we find intrinsic scatters in the  fits of 0.055, 0.067, 0.055 (dex of $V$ km $s^{-1}$) at $\langle z \rangle$=0.37, 0.62, 0.96, which are comparable to that seen in local TF relations (i.e., $\sim$0.049 in \citet{pizagn2005}). Our analysis therefore represents a significant improvement on the scatter seen in earlier intermediate-redshift studies, for example an improvement of a factor of 2-3 over the study of \citet{consel2005}. We also achieve a scatter less than that found in the $M_{\ast}$-$S_{0.5}$ relation of \citet{kassin2007} (between 0.08 and 0.11 dex), despite the fact that they introduce an additional dispersion term, $S_{0.5}$, which significantly tightens the relation from the log $V$ - $M_{\ast}$ relation. 

Table~\ref{table_smtfr} shows little room for evolution in the relation. Fitting a straight line through the zero-points between redshift bins, we detect a modest but statistically insignificant trend for a larger stellar mass at fixed velocity at lower redshift: $\Delta M_{\ast}$$\sim$$0.037 \pm 0.065$ dex (1$\sigma$) from $\langle z \rangle$$\sim$1.0 to 0.3.  Fitting a straight line through zero-points between redshift bins in the \citet{consel2005} study, we see a $\Delta M_{\ast}$$\sim$$0.07 \pm 0.19$ dex (1$\sigma$) from $\langle z \rangle$$\sim$1.0 to 0.3, consistent with our result, yet more uncertain. Interestingly, our 1-sigma limit for evolution is consistent with the modest $\Delta \log{M_{\ast}/M_{\odot}}$ predicted by \citet{portin2007} ($\Delta M_{\ast} \sim0.1$), \citet{somerv2008a} (10\% decrease in $V$ at fixed $M_{\ast}$ with time), and only consistent with \citet{dutton2011a} ($\Delta M_{\ast} \sim0.2$) over a similar redshift range at the 2-sigma level. Importantly, this result is robust to the inclusion of the local data in both the {\it enclosed} and {\it total} relations. We discuss the implications of this agreement further in \S \ref{sec:TFR}. 

We note that the scatter in the {\it total} stellar mass TF relations is increased somewhat compared to that in our preferred {\it enclosed} relations, most likely due to the effect of extrapolated velocity measurements. We trace spectroscopic emission beyond $r_D$ on $\sim$60\% of our disks as opposed to $\sim$90\% beyond $r_{2.2}$. 

\subsection{The $B$-band magnitude TF relation}\label{sec:bmtfr}

\begin{figure*}[h]
   \centering
 \includegraphics[width=5.3in]{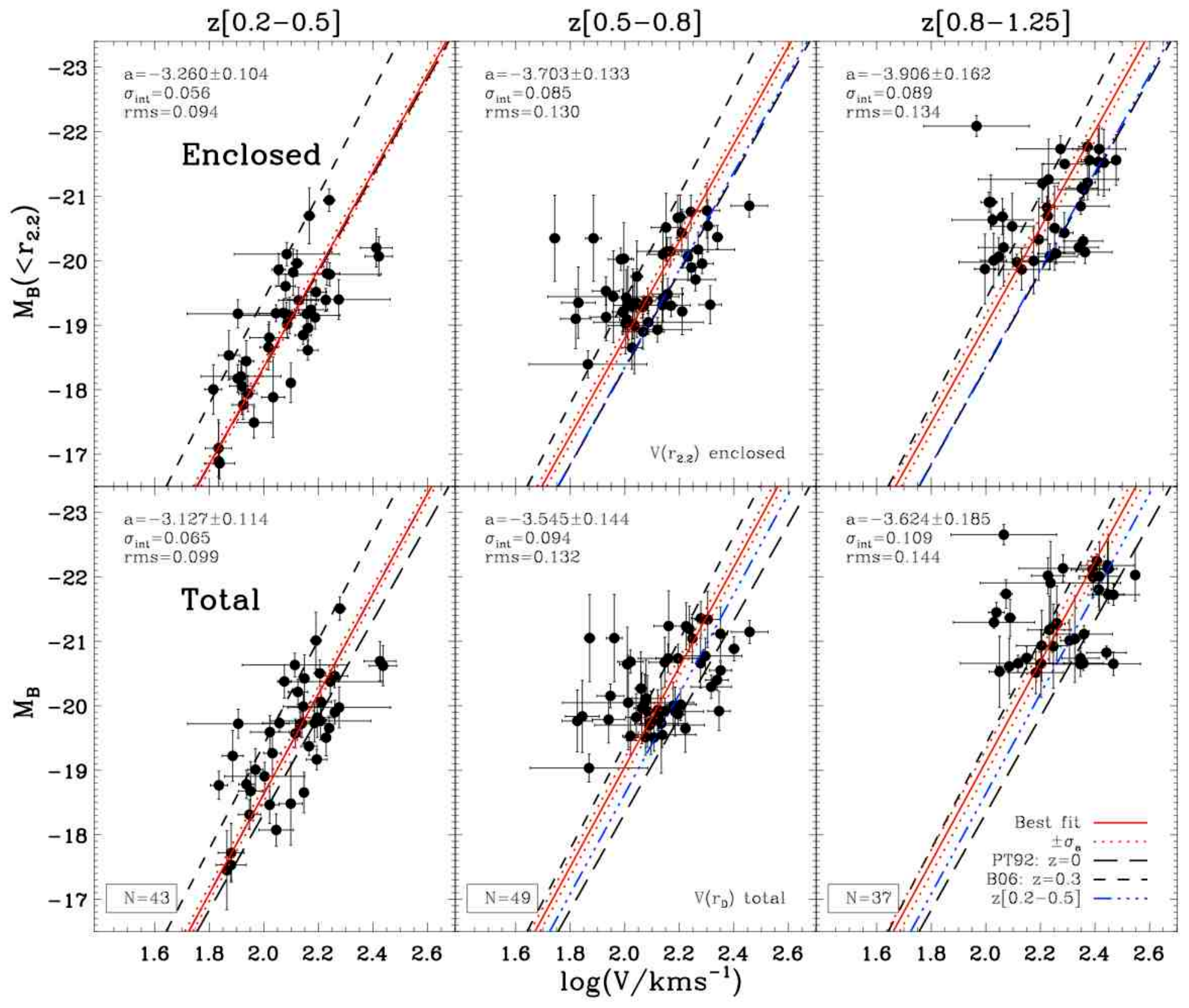}    
      \caption{As Fig.~\ref{fig:clas_tf} but for the absolute $B$-band magnitude ($M_B$) Tully-Fisher relation. For comparison purposes, we show the local relation of \citet{pierce1992} and the $z\sim\langle0.3\rangle$ relation of \citet{bamfor2006} as dashed lines. Other lines are as in Fig. ~\ref{fig:clas_tf}.} 
   \label{fig:bmag}
\end{figure*}

\begin{table*}[h]
 \begin{tiny}
 \caption{Absolute $B$-band Magnitude Tully-Fisher Relations}
\begin{tabular}{ccc}
\hline
 \end{tabular}
\begin{tabular}{lllllllllll}
\hline
z range & $\langle z \rangle$ &  N &  $a$\footnote{best-fit y-intercept in mag assuming scatter in $V/km s^{-1}$ dex} & $b$\footnote{slope assuming scatter in $V/km s^{-1}$ dex} & $\sigma_{int,V}$\footnote{internal scatter in $V/km s^{-1}$ dex} & median $\sigma_V$\footnote{median velocity error in $V/km s^{-1}$ dex} & $rms_V$\footnote{total scatter in$V/km s^{-1}$ dex} & $\sigma_{int,M}$\footnote{internal scatter in mag} & median $\sigma_M$\footnote{median $B$-band magnitude error} & $rms_M$\footnote{total scatter in mag} \\
[0.5ex]
\hline
Enclosed: $M_B$($r_{2.2}$) vs  V($r_{2.2}$): \\
[0.5ex]
\hline
\textbf{0.2$<$z$\le$1.3} & \textbf{0.64} & \textbf{129} & \textbf{-3.589$\pm$0.893} & \textbf{-7.546$\pm$0.578} & \textbf{0.081} & \textbf{0.022} & \textbf{0.127} & \textbf{0.612} & \textbf{0.291} & \textbf{0.956}\\
0.2$<$z$\le$0.5 & 0.37 & 43 & -3.260$\pm$0.104 & " fixed & 0.056 & 0.035 & 0.094 & 0.425 & 0.245 & 0.711 \\
0.5$<$z$\le$0.8 & 0.62 & 49 & -3.703$\pm$0.133 & " fixed & 0.085 & 0.045 & 0.130 & 0.641 & 0.364 & 0.979 \\
0.8$<$z$\le$1.3 & 0.96 & 37 & -3.906$\pm$0.162 & " fixed & 0.089 & 0.062 & 0.134 & 0.670 & 0.299 & 1.011 \\
[0.5ex]
\hline
Total:  $M_B$ vs  V($r_D$) \\
[0.5ex]
\hline
\textbf{0.2$<$z$\le$1.3} & \textbf{0.64} & \textbf{129} & \textbf{-3.413$\pm$1.037} & \textbf{-7.754$\pm$0.657} & \textbf{0.091} & \textbf{0.023} & \textbf{0.130} & \textbf{0.706} & \textbf{0.250} & \textbf{1.008}\\
0.2$<$z$\le$0.5 & 0.37 & 43 & -3.127$\pm$0.114 & " fixed & 0.065 & 0.037 & 0.099 & 0.505 & 0.258 & 0.771 \\
0.5$<$z$\le$0.8 & 0.62 & 49 & -3.545$\pm$0.144 & " fixed & 0.094 & 0.057 & 0.132 & 0.731 & 0.364 & 1.027 \\
0.8$<$z$\le$1.3 & 0.96 & 37 & -3.624$\pm$0.185 & " fixed & 0.109 & 0.064 & 0.144 & 0.845 & 0.272 & 1.114 \\
[0.5ex]
\hline
\hline
\end{tabular}
\label{table_bmag}
\end{tiny}
\end{table*}

We undertake a similar analysis to that described in \S \ref{sec:tfrs} for the absolute $B$ magnitude ($M_B$) TF relation, where $M_{B}$ measurements come from the best-fit SEDs (described in \S \ref{sec:masses}), and are aperture corrected in the same way as the stellar mass estimates. The TF relations are shown in Fig.~\ref{fig:bmag} for both total and enclosed luminosities and the results are listed in Table~\ref{table_bmag}. Any difference in the redshift-dependent trends compared to that for stellar mass relation would indicate changes in the star formation rate per unit stellar mass. Some workers have claimed to see evolution in the $M_B$-TF relation \citep{weiner2006b,fernan2010} and we aim to verify or otherwise these trends with our improved dataset. As before, we use the local relations of \citet{pierce1992} and the $\langle z \rangle$$\sim$0.3 study of \citet{bamfor2006} as comparison datasets. 

Once again, the intrinsic scatter around our redshift-dependent $M_B$-TF relations, 0.424, 0.641, 0.670 in mag, at $\langle z \rangle$=0.37, 0.62, 0.96, are comparable to those seen in the local relations ($\sim$0.4 mag in \citet{pierce1992}, and 0.3-0.5 in \citet{verhei2001}), and we note a considerable improvement over previous intermediate-redshift studies. The $\langle z \rangle$$\sim$0.3 study of \citet{bamfor2006} and the $\langle z \rangle$$\sim$0.85 study of \citet{chiu2007} have scatters twice as large ($\sim$0.9 in mag), and those of \citet{weiner2006a} to z$\sim$1.2 and \citet{fernan2010} to z$\sim$1.4 have scatters $\simeq$2-3 times as large ($\sim$1.5 mag and $\sim$1.2 mag) respectively. 

Even though the $M_B$-TF relation is not as tight as our $M_{\ast}$-TF relation, there is evidence for a stronger evolution in the $M_B$-TF relation than in the $M_{\ast}$-TF relation, as expected from the well-established increase in disk star formation rate to $z \sim 1$ \citep{bundy2005}. Fitting a linear regression through the zero-points between redshift bins of the enclosed $M_B$-TF relation we find $\Delta M_{B}$$\sim$$0.85 \pm 0.28$  mag evolution in the relation from $\langle z \rangle$$\sim$1.0 to 0.3. We can check if this result is affected by a Malmquist bias (given the distribution of luminosities is significantly different between redshift intervals), by comparing subsets with similar luminosities and stellar masses, and the trends do not substantially change. \citet{weiner2006a} find a consistent trend of $\sim$1.0-1.5 mag evolution from a similar redshift range of $\langle z \rangle$$\sim$1.2 to 0.4, but with more uncertainty. Our results are consistent with the evolution shown in the models of \citet{portin2007}: $\Delta M_{B} \sim 0.85$ from $z\sim1$. We can understand the different evolutionary trends in the $M_B$-TF and $M_{\ast}$-TF relations by examining the redshift-dependent correlation between $M_{B}$ and $M_{\ast}$. To first order, as expected, the difference is explained by the increase in the $B$ band luminosity per unit stellar mass with redshift  \citep{lilly1996,madau1996}.  

\section{Interpreting the Tully-Fisher Relation}
\label{sec:TFR}

We now seek a physical interpretation of the results presented in \S \ref{sec:reses} in the context of current models of disk galaxy assembly. First we discuss various procedures for estimating {\it dynamical masses} from our rotation curve data (\S \ref{sec:dyn_est}). We then derive estimates of the total {\it baryonic mass} (\S \ref{sec:bar_est}). We combine the two estimates to evaluate the relative roles of baryons and dark matter out to the observable radii probed with our deep exposures (\S \ref{sec:bar_vs_dyn}). Although there are considerable uncertainties in what follows, our intent at this stage is to illustrate the possibilities that will arise when gas masses can be determined for samples such as ours so that the total baryonic components would be accurately measured and their role in the Tully-Fisher relation established.

\subsection{Dynamical Mass Estimates}\label{sec:dyn_est}

The physical basis of our interest in the Tully-Fisher relation is that the dynamical mass is strongly correlated with the luminous and stellar mass components of galaxies, and by analyzing empirical constraints, we can gain an understanding of the relative assembly histories of dark and baryonic matter in galaxies. We thus seek to use our data to estimate both the dynamical masses (i.e., the total mass, including dark and baryonic) as well as that of the stars and gas. Previous studies of this nature \citep[ all low-redshift galaxies]{pizagn2005, gnedin2007, willia2010} derived dynamical masses from kinematic data that probe sufficiently far in radius to detect the dark halo by revealing a deficit of baryons when dynamic masses are compared to stellar masses.

However, our method of using emission line velocities to estimate the mass within a given radius, such as $r_{2.2}$, depends sensitively on the assumed shape of the underlying gravitational potential, and hence the distribution of mass throughout the disk. For a given ellipsoid potential, the velocity can be most simply approximated as:
\begin{equation}
V_{c}(r)^2 \approx \xi \frac{G M(r)}{r},
\label{eq:ellip_pot}
\end{equation}
where $\xi$=1 in the case of spherical symmetry. Assuming a {\it spherical potential} will likely overestimate the disk mass unless a spherical dark matter halo is dominant within the relevant radius. As such it supplies an effective upper limit for a given mass of an ellipsoid calculated from the observed circular velocity. Traditionally, dynamical disk masses have been calculated with an {\it exponential `Freeman' potential}, solved with modified Bessel functions by assuming a constant mass-to-light ratio and an infinitely-thin disk of infinite size \citep{freema1970}. This ignores the presence of the bulge and halo, known to be important even at the scales considered here \citep[e.g.,][]{trott10,dutton11}. Therefore, as an alternative method of estimating a lower limit, we adopt an {\it oblate potential}, characterized by a flattening factor $q$, which is the ratio of the scale length normal to the disk over the scale length of the disk. As shown in \citet{binney1987}, the velocity for an oblate sphere can then be considered as:
\begin{equation}
V_{c}(r)^2 \approx 4 \pi G q  \int_0^r \frac{\rho(m^2) m^2 dm}{ \sqrt{r^2 - m^2(1 - q^2)}},
\label{ob:oblate_pot}
\end{equation}
where $m^2 = r^2 + r_s^2$, and $\rho$ is the assumed density function. The exact shape of the potential will depend on the relative contribution of luminous and dark matter, as well as on the triaxial shape of each component. Although halos are believed to be prolate on large scales, their shape is less clear at the scales considered here. Lensing and dynamical studies of individual systems \citep{dutton11} suggest that they may be considerably rounder. Furthermore, the presence of a bulge generally implies that the stellar distribution is significantly less flat than that of a pure disk. We adopt $q=0.4$ and an exponential density function as a representative maximum oblateness, equivalent to $\xi \approx 0.752$ for Eq.\ref{eq:ellip_pot}. If we were to use a de Vaucouleurs profile (S\'{e}rsic profile where $n$ = 4) instead of an exponential density profile, $\xi$ would be $\approx$ 0.833, resulting in a less than 10\% change in the dynamical mass calculation. In the following we will consider the $q=1$ spherical case and the $q=0.4$ oblate case as bracketing the shape of the total potential. Additional systematic uncertainties include the effects of non-streaming motions, warps and non-gravitational forces, as discussed in the well-established literature on the interpretation of local rotation curves \citep[see][and references therein]{binney1987}. Finally, we consider possible biases arising from slit spectroscopy and its maximum effect propagated to our dynamical mass estimates in the following section.

\subsection{Slit-effect correction}\label{sect:slit_corr}

Recent progress with integral field unit (IFU) spectrographs has illustrated some limitations of traditional long-slit and multi-slit techniques in determining the internal dynamics of intermediate redshift galaxies. This slit-effect is similar to beam smearing in radio astronomy, where the range of velocities from incoming light are averaged over the width of the slit, resulting in a broadening of Doppler shifted lines and an average reduction of the rotational velocity.   The magnitude of this effect has been considered in detail by \citet{kapfer2006} and used by \citet{flores2006} to compare IFU-derived rotational velocities to those determined with a multi-slit instrument. \citet{kapfer2006} systematically investigated the effects of various slit widths in combination with inclination, spatial binning and position angle offsets on measured disk velocities using N-body/SPH simulations. We can use these results to consider the effect our slit geometry might have in distorting our velocities taking into account the galaxy sizes and shapes relative to the DEIMOS slits. Kapferer et al. find no systematic bias due to binning and position angle offset (beyond the correction already made, \S \ref{sec:fit}). We can, however, use Kapferer et al.'s results to calculate an upper-limit approximation of the correction to the velocity $V_{2.2}$ for the effect of the slit width relative to scale radius ($r_s$) of each galaxy according to its axis ratio ($b/a$) and for a given inclination $i$. Derived from Figs 9 \& 10 of \citet{kapfer2006}, the correction is:
\begin{equation}
V_{corr} = V_{obs} + V_{beam} \frac{1}{(b/a)} \frac{r_{slit}}{r_s} \sin{i},
\label{eq:beam}
\end{equation}
where $V_{beam} \sim 20~\mathrm{km~s^{-1}}$. We find correlations in the correction with respect to scale radius and inclination, but none with mass, redshift, position angle or observed velocity. The correction added to $V_{2.2}$ ranges from 2 km $\mathrm{s^{-1}}$ to 52 km $\mathrm{s^{-1}}$, with a mean of 20 km $\mathrm{s^{-1}}$. These corrections can be found in our catalog. 

Because of the imprecise nature of these corrections, arising from the fact that the Kapferer et al. result assumes a symmetric Gaussian to the spectral line profile (while we fit for two half-Gaussians to account for much of the blending between the seeing and dispersion), the Kapferer et al. -based correction remains an upper-limit. Thus we did not include them in our precisely measured Tully-Fisher relations in \S 4. However, we will apply them to our dynamical mass calculation in order to not bias our estimates in a way that may over-estimate the dominance of the stellar mass compared to the dark matter. The difference with and without the slit-effect correction can be seen in Fig.~\ref{fig:bmf}. Because of the imprecision of the analytical formula derived here, we strongly advise to not apply such a formula beyond its tested range.

\subsection{Baryonic mass estimates}
\label{sec:bar_est}

In order to examine the redshift-dependent fraction of baryonic mass within $r_{2.2}$ we need to obtain an estimate of the total baryonic mass.  In addition to stellar masses, discussed in~\S~\ref{sec:masses}, we need to account for the presence of gas.

Accurate gas masses are not yet available for intermediate redshift galaxies, although CO-derived masses have begun to appear for some systems at z$>$1 with, e.g., the PdBI interferometer \citep{tacconi2010, daddi2010}. Nonetheless the situation will improve significantly through upcoming facilities such as ALMA, MeerKAT, and eventually the Square Kilometer Array. Although what follows is somewhat speculative, it provides a reasonable illustration of what will soon be possible. To make progress, we estimated gas masses ($M_g$) for our sample using the local stellar-to-gas mass ($M_{\ast}$-to-$M_{g}$) ratio as a function of $M_{\ast}$, recently parameterized by \citet{peeple2010} based on H I measures from The H I Nearby Galaxy Survey (THINGS) and helium-corrected, CO-derived $\mathrm{H}_2$ masses from the HERA CO-Line EXtragalactic Survey (HERACLES) and the Berkeley-Illinois-Maryland Association Survey of Nearby Galaxies (BIMA SONG) \citep{leroy2008}.

According to the parameterization by \citet{peeple2010}:
\begin{equation}
\frac{M_g}{M_{\ast}} = K_{f}M^{-\gamma}_{\ast}
\label{equ:m_g_vs_m_s}
\end{equation}
where $K_{f}$ = 316228, $\gamma$=0.57, and $M_{\ast}$ is measured in units of solar masses. 

Until precision gas masses become available we cannot be certain that the \citet{peeple2010} formalism can be applied in this manner at intermediate redshift. However, locally-measured gas-to-stellar mass ratios will underestimate the gas mass for intermediate-redshift galaxies since many stars have subsequently formed. To correct for this, we consider the observed evolution in the specific star-formation rate (sSFR) for blue galaxies to z$\sim$2 as measured by \citet{oliver2010}. To determine the correction for each galaxy, we integrate the best-fit sSFR relation out to the relevant redshift. \citet{oliver2010} find:
\begin{equation}
\mathrm{sSFR} = X(1+z)^\alpha
\label{eq:sSFR}
\end{equation}
where $\log_{10}{X/Gyr^{-1}} = -1.36\pm0.41$ and $\alpha =-3.4\pm0.3$. We temper this correction by the gas recycling rate \citep[e.g.,][]{kennic1994,madau1998,cole2001}, estimated to be $\sim$40\% for a \citet{chabri2003} initial mass function from z$\sim$1 to present. This scenario is not strictly a closed-box model, where a galaxy sits in a huge reservoir of gas and simply converts that gas to stars over time. For one, we include the substantial Chabrier recycling rate consistent with the IMF we assume in our stellar mass estimates. The scenario we present here is also consistent with roughly equal inflow and outflow since $z\sim1$. The fraction of feedback driven material that exceeds the escape velocity of the galaxy's potential well, and also the typical redshifts at which the filaments that feed galaxies evaporate, are still largely unknown. The gas mass estimates we add to our stellar masses are on average 30\% of the total stellar mass estimates, and only exceed that of the stellar mass in a few of the lowest mass objects. So we claim that our gas mass estimates are not the main drivers of our conclusions, and each galaxy's stellar mass and redshift essentially determine the gas mass estimate for this scenario.

\subsection{Comparison of Baryonic and Dynamical Masses}
\label{sec:bar_vs_dyn}

\begin{figure*}
   \centering
   \includegraphics[width=3in]{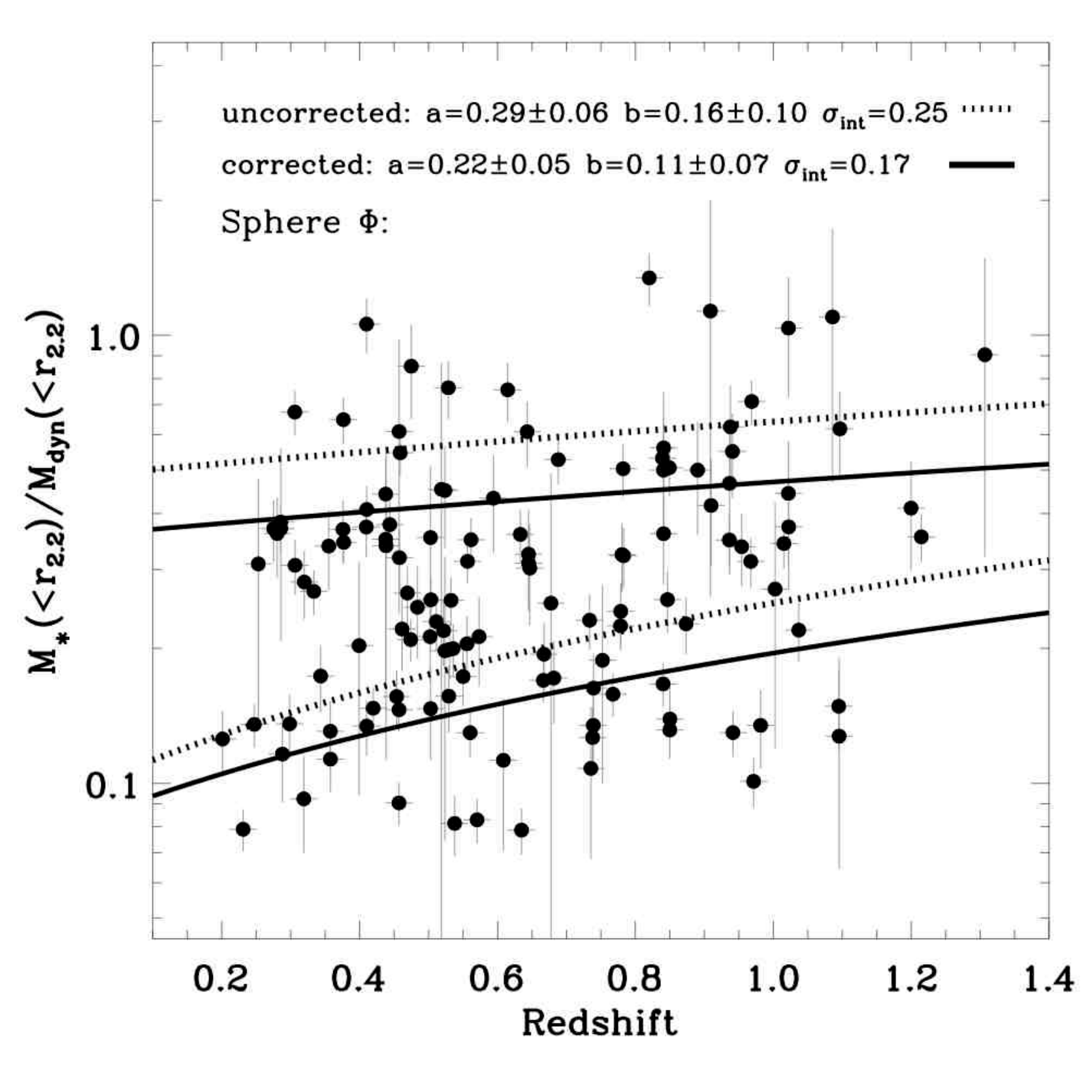}
   \includegraphics[width=3in]{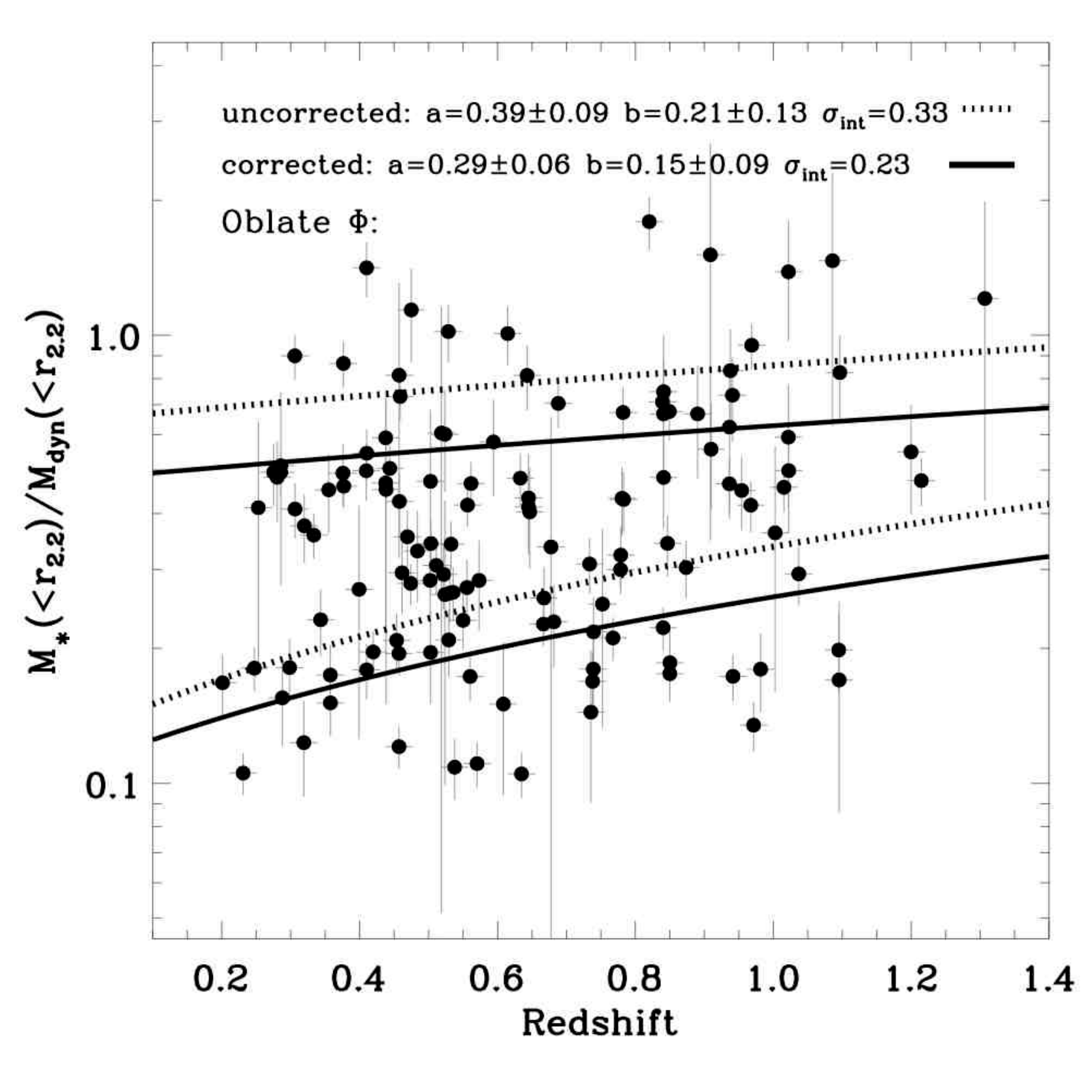}
    \includegraphics[width=3in]{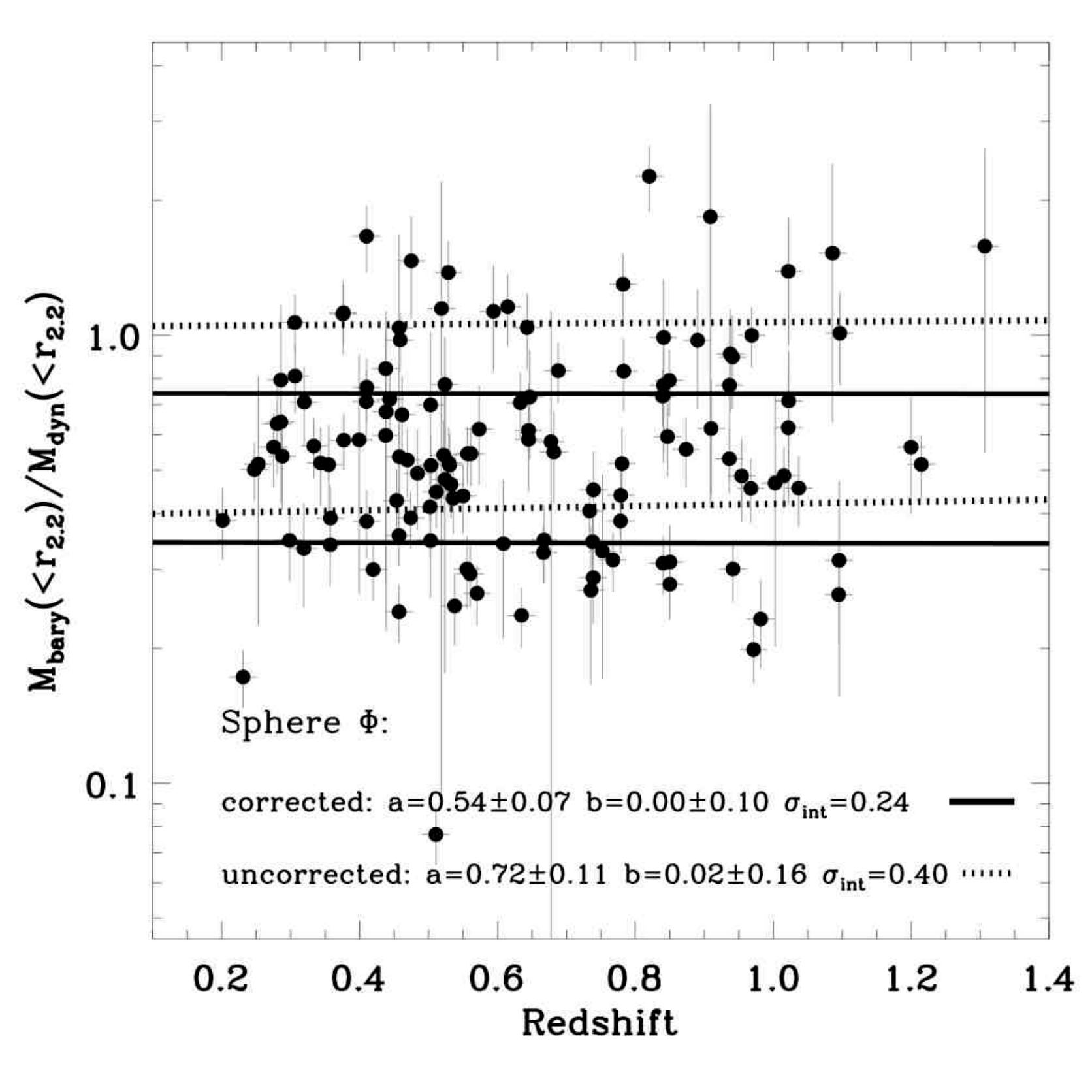}
    \includegraphics[width=3in]{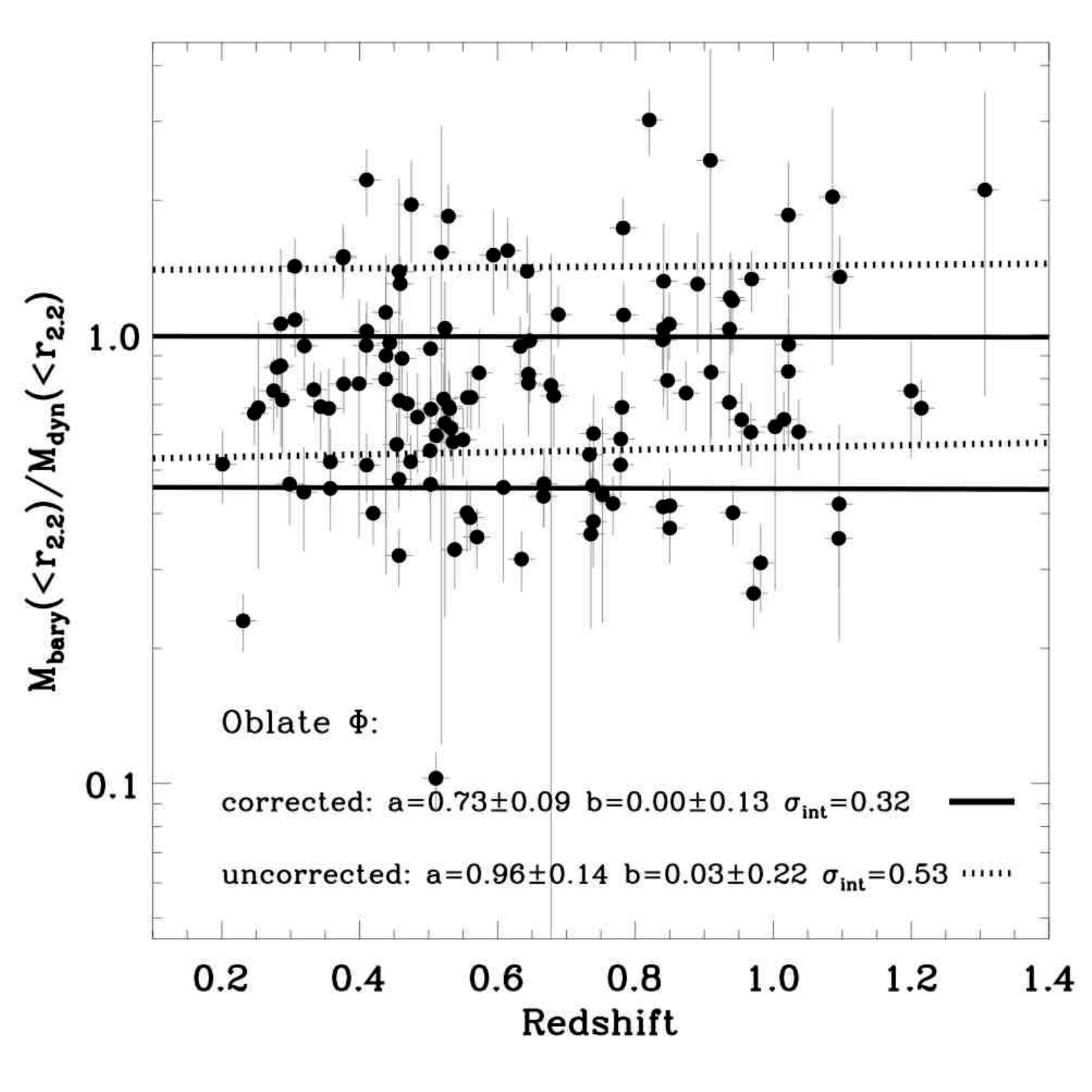}
      \caption{Top panels: stellar-to-dynamical mass ratio within our fiducial $r_{2.2}$ aperture as a function of redshift, assuming a spherical potential (left) and an oblate potential (right). Curves indicate 3$\sigma$ variations around the best-fit linear relation, with slope $b$ and intercept $a$. The solid lines are derived from dynamical masses including the slit-effect correction, and the dotted lines are derived from the uncorrected dynamical masses. The scatter around the best fit relation is also given in each plot. Bottom panels: same as the top panels for baryonic (stellar plus gas) mass.}
   \label{fig:bmf}
\end{figure*}

We finally turn to our comparison of the dynamical mass with the stellar and baryonic mass. The results of this comparison are given in Figure~\ref{fig:bmf}. The top panels show the stellar-to-dynamical mass ratio for a spherical and oblate potential.  Across our redshift range we find our stellar-to-dynamical mass fractions are $\sim0.3$ for a spherical potential and $\sim0.4$ for an oblate potential, with considerable scatter. The dynamical mass estimates for the points plotted include the slit-effect correction (\S \ref{sect:slit_corr}), representing an upper-limit for the dynamical masses, and thus a lower limit for the stellar-to-dynamic mass fraction. For comparison in the plots of Figure~\ref{fig:bmf}, the dotted lines are derived from the uncorrected data, and represent a lower limit to the dynamical masses, given an assumed potential shape (spheroid or oblate in our example). The points without slit-effect corrections are not plotted, and just their 3$\sigma$ best-fit contours are plotted for simplicity. In a few cases, the stellar-to-dynamical fractions are as high as unity, suggesting that baryons play a significant role in driving the TF relation. A dominant baryon fraction is broadly consistent with earlier results by \citet{consel2005}, \citet{gnedin2007}, \citet{dutton2009} and \citet{dutton2010b}, especially considering the many uncertainties involved in both mass estimates, including the stellar initial mass function (IMF).  

The hypothesis that baryonic mass primarily governs the slow redshift-dependent trends in our observed TF relations is supported further when we attempt to add estimates for the missing gaseous components to our stellar masses. We then find that our baryonic mass estimates within $r_{2.2}$ approach those determined dynamically, with no redshift dependence (bottom panels of Figure~\ref{fig:bmf}). The average baryonic-to-dynamical mass ratio is $\sim50$\% and $\sim70$ \% respectively for spherical and oblate potentials (including slit-effect corrections). It must be remembered this baryonic estimate does not include any ionized gas. 

Next we explore the radially dependent profile of the dynamical mass as calculated from the rotation curve, to compare to the baryonic component mass profile. At each tenth of a scale radius along the profile, starting at 1 scale radius, we compute the dynamical mass given the best modeled velocity at that radius in the rotation curve, and compare that to the enclosed baryonic mass at that radius. The baryonic mass is assumed to follow the stellar mass profile, which we approximate to follow the distribution of light found in the reddest HST filter (F850LP), in an aperture stellar mass estimate, the method of which is described in detail in \S\ref{sec:masses}. The enclosed light at each tenth of a scale radius is compared to the total light, stepping out in elliptical apertures from the photometry, and the aperture mass is calculated from these ratios.

Although past studies of rotation curves and the TF relation were often motivated as a means of tracing the influence of dark matter on the baryonic assembly of galaxies, \citep[e.g.,][]{rubin1980,vogt1997,pizagn2005,consel2005}, it has been suggested in local TF studies that the role of dark matter is minimal in the optical disk \citep[e.g.,][]{courte1999,paluna2000,bellde2001,kassin2006}.  Figure \ref{fig:aperture} shows that, irrespective of the aperture $f$, defined in terms of the scale radius (i.e. $f$=2.2 for $r_{2.2}$), out to the furthest observable extent in our disks, the mean baryonic-to-dynamical fraction is consistent with unity when using an oblate potential for our dynamical mass estimates. In contrast, for a spherical potential, we find our results permit an equal contribution from dark and baryonic matter by $r_{2.2}$, consistent as we noted above with \citet{portin2007} and \citet{dutton2011b}. 

We also show the baryonic-to-dynamical fraction with respect to aperture for a \citet{freema1970} potential. The unphysical and sharp rise above unity for small radii shows that the dynamical mass is likely underestimated for the central-most part of the galaxy using this potential. If used in a maximal disk fit, which would re-normalize the sharply rising peak down to unity, the fit would likely result in an underestimate of the overall dynamical contribution of the baryons for what is nominally the scenario where baryons are maximally contributing to the potential. This gives us some indication that the Freeman potential, which assumes a constant mass-to-light ratio for an infinitely-thin exponential disk,  may not be the best approximation in maximal/minimal fit applications for a disk of a finite thickness and size.

We thus conclude for the scenario presented here that the contribution by baryons to the total mass within the radial range probed by observations is between 50\% and 100\%. Baryons appear to be the most important component within 2.2 scale radii, and perhaps to larger radii, depending on the uncertain conversion from circular velocity to stellar mass. In addition, such a high fraction of baryons is expected to influence significantly the overall profile of the dark matter halo at these scales through gravitational interactions \citep[see, e.g.,][and references therein
for a discussion of this topic in the local universe]{noordermeer07}. Therefore even if dark matter is present in equal amount, it will be tightly coupled to the baryonic content.

\begin{figure}
   \centering
   \includegraphics[width=3.5in]{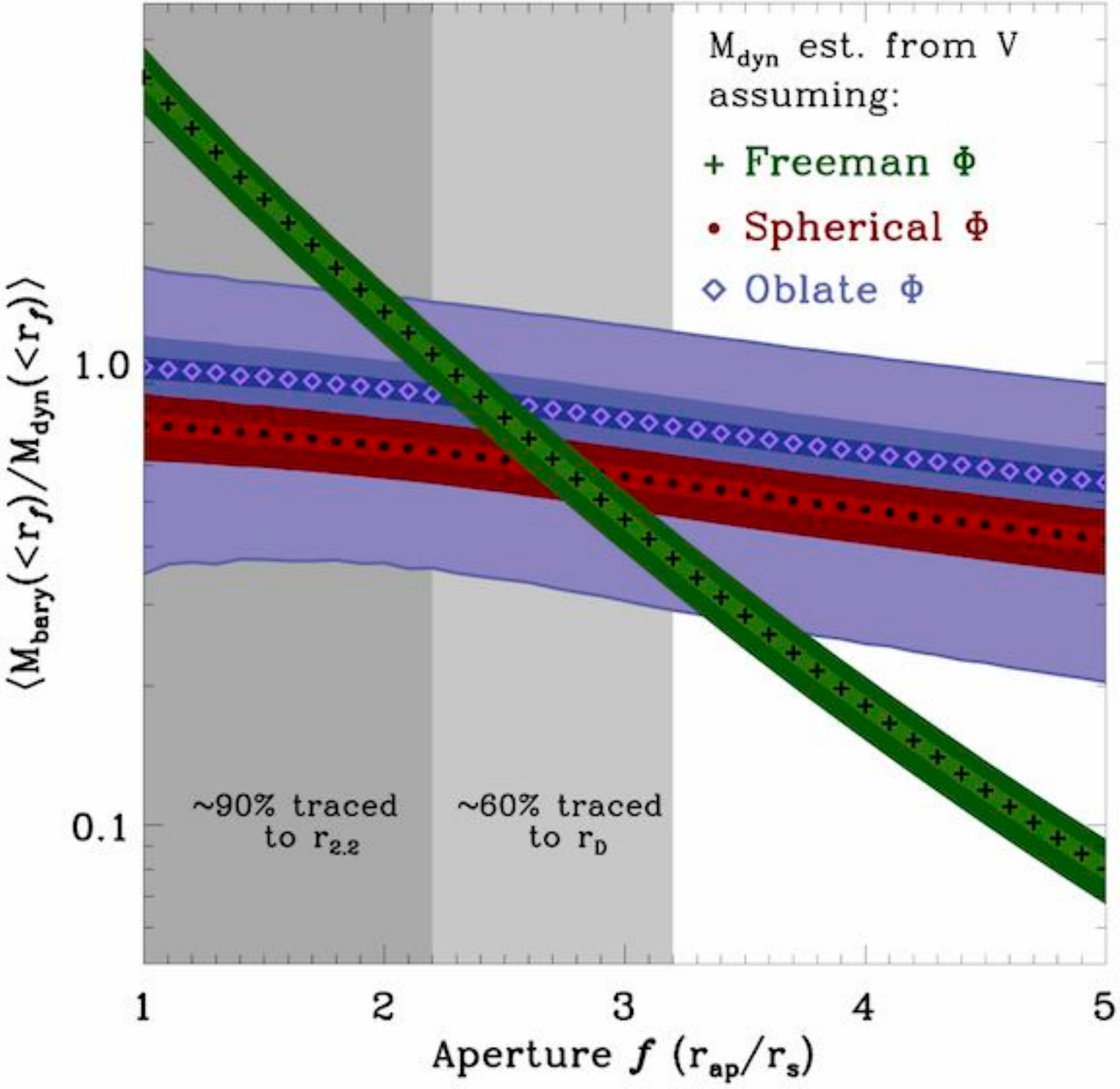}
      \caption{Estimates of the mean baryonic-to-dynamical mass fraction over all redshifts as a function of relative aperture ($f=r/r_s$). Shaded regions show the percentage of disks traced to $f$=2.2 and 3.2 ($r_D$); our Keck survey adequately samples the rotational velocities to $f\simeq$3.5. Different curves relate to different determinations of the dynamic mass (see text for details) and the tight contours show the 1 and 3 times the standard error on the mean. Light shading around the results assuming an oblate potential indicate the aperture-dependent region containing 68\% of the individual fractions for our disks. These regions are not shown for the other two potentials, but are of similar size to that of the oblate potential. }
   \label{fig:aperture}
\end{figure}

\section{Summary \& Discussion} \label{disc}

Using spectra with extended 6-8 hour integration times secured with the DEIMOS instrument on the Keck II telescope we have measured 129 rotation curves for galaxies selected with $z_{AB}$$<$22.5 in the redshift range 0.2$<$$z$$<$1.3 in the two GOODS fields. Using associated HST imaging data, we derive modeled velocities corrected for the effects of inclination and seeing to a fiducial radius, $r_{2.2}$, corresponding to 2.2 times the disk scale length. As 90\% of our rotation curves flatten within this radius, this gives us a highly reliable set of kinematic data which we use to construct the stellar mass ($M_{\ast}$) and B-band ($M_B$) Tully-Fisher (TF) relations and their evolution over the past 8 Gyr. We find the following results.

\begin{enumerate}

\item We demonstrate a significantly reduced scatter around our intermediate redshift TF relations compared to that seen in earlier studies. The scatter around the $M_{\ast}$-TF relation is typically 0.2 dex of $M_{\ast}/M_{\odot}$, which is 2-3 times less than earlier work, comparable to that determined locally and less than that achieved by \citet{kassin2007} who introduced an additional dispersion term in order to achieve a tight relation. The scatter around our $M_B$-TF relation is typically 0.4-0.7 mag, again comparable to that in local relations and a factor of 2-3 improvement over earlier work at intermediate redshift. In addition to demonstrating the validity of our observational approach and our improved modeling techniques, our results clearly show the TF relation is well-established at redshift $z\simeq$1. 

\item We find a modest but statistically-insignificant evolution in the $M_{\ast}$-TF relation with redshift corresponding to a growth in stellar mass at fixed velocity of $\Delta\,M_{\ast}$ = 0.04 $\pm$0.07 dex with cosmic time from a redshift $\langle$$z$$\rangle$$\simeq$1.0 to 0.3. This is consistent with the results of recent numerical and semi-analytic simulations \citep{portin2007,somerv2008a,dutton2011a}.

\item More pronounced evolution is seen in the $M_B$-TF relation corresponding to a decline in luminosity of 0.85 $\pm$ 0.28 mag from $\langle$$z$$\rangle$$\simeq$1.0 to 0.3, again consistent with predictions from \citet{portin2007}. The origin of this evolution can be understood by examining the mass-dependent correlation between $M_{\ast}$ and $M_B$ as a function of redshift. Together with the larger scatter seen in the $M_B$-TF relation than in that based on stellar mass, this demonstrates that the $M_{\ast}$-TF relation is the more fundamental descriptor of disk galaxy assembly.

\item We finally use our data to illustrate the potential of comparing dynamical and baryonic masses to better understand the contributions that they matter make in the $M_{\ast}$-TF relation. Although we are uniquely placed to explore our extended rotation curves and carefully-derived stellar masses, our gas mass estimates are clearly approximate at this stage. We estimate dynamical masses using both spherical and oblate potentials which we expect bracket the likely values. We compute baryonic masses by estimating the additional mass in cold gas.  We find that baryons may contribute between 50-100\% of the total mass within the radii at which we can confidently observe dynamics of the gas in our disk sample. Such a high fraction of baryons influence significantly the overall profile of the dark matter halo. Even if dark matter is present in equal amount, it will be tightly coupled to the baryonic content. 

\end{enumerate}

\acknowledgments

SHM thanks the Rhodes Trust, the Oxford Astrophysics Department, New College, Oxford, and the California Institute of Technology for supporting her work. MS and RSE acknowledge financial support from the Royal Society. We thank S. Moran, N. Miller, G. Mikelsons, and K. Chiu for considerable initial assistance with the organization of this dataset. We thank C. Peng for supplying us with \textsc{GALFIT} 3.0 and acknowledge useful discussions with A. Benson, A. Brooks, A. Bunker, M. Bureau, M. Cappellari, R. Davies, H. Flores, F. Hammer, S. Kassin, L. Miller, and M. Puech. The referee, B. Weiner, is acknowledged for his constructive comments in the improvement of this paper. The spectroscopic data was secured with the W.M. Keck Observatory on Mauna Kea. We thank the observatory staff for their dedication and support. The authors recognize and acknowledge the cultural role and reverence that the summit of Mauna Kea has always had with the indigenous Hawaiian community. We are most fortunate to have the opportunity to conduct observations from this mountain.

\bibliography{mybib}{}
\bibliographystyle{apj.bst}

\clearpage

\begin{turnpage}
\begin{tiny}
\begin{table}
\caption{Table of Measurements: first 20 entires}
\label{table:measurements}
\begin{tabular}{llllllllllllllll}
\hline
\hline
RA & Dec & $z$ & $m_z$ & $PA_{slit}$\footnote{in degrees} & $PA_{off}$\footnote{in radians} & $\sin{(i)}$\footnote{sin of inclination} & $r_{2.2}$\footnote{2.2 $\times$ scale radius in kpc} & $M_{*}(r_{2.2})\footnote{enclosed stellar mass in log $M_{\ast}/M_{\odot}$ dex}$ & $M_{B}\footnote{total absolute B-magnitude in mags}$ & $M_{K_s}\footnote{total absolute K-magnitude in mags}$ & $V_{2.2}$\footnote{best modeled velocity at $r_{2.2}$ in $km s^{-1}$} & $V_{slit}$\footnote{estimate of upper limit of velocity broadening of slit, in $km s^{-1}$} & Lower $M_{dyn}$\footnote{enclosed dynamical mass in log $M_{\ast}/M_{\odot}$ dex, lower limit (without slit effects correction, assuming an oblated potential where q=0.4)}  & Upper $M_{dyn}$\footnote{enclosed dynamical mass in log $M_{\ast}/M_{\odot}$ dex, upper limit (with slit effects correction, assuming a spherical potential)}\\ [0.5ex]
\hline

       189.28400 & 62.204340 & 0.59 & 22.22 & 36.10 & 0.00 & 0.97 & 2.87$\pm$0.21 & 9.44$\pm$0.17 & -19.83$\pm$-0.56 & -22.76$\pm$-0.64 & 67.55$\pm$10.02 & 29.92 & 9.36$\pm$0.11 & 9.80$\pm$0.11 \\
       189.34309 & 62.196030 & 0.53 & 21.97 & 81.30 & 0.73 & 0.96 & 5.26$\pm$0.19 & 9.11$\pm$0.14 & -19.76$\pm$-0.48 & -22.41$\pm$-0.54 & 66.07$\pm$6.31 & 16.26 & 9.60$\pm$0.07 & 9.92$\pm$0.07 \\
       189.32520 & 62.213470 & 0.91 & 22.17 & 36.70 & 0.02 & 0.96 & 10.73$\pm$0.54 & 10.77$\pm$0.03 & -20.82$\pm$-0.10 & -24.97$\pm$-0.10 & 227.47$\pm$39.66 & 10.26 & 10.99$\pm$0.13 & 11.15$\pm$0.13 \\
       189.38380 & 62.212980 & 1.02 & 21.22 & 76.40 & 0.02 & 0.79 & 13.84$\pm$0.24 & 11.40$\pm$0.12 & -22.17$\pm$-0.36 & -26.23$\pm$-0.43 & 260.36$\pm$51.88 & 12.74 & 11.21$\pm$0.14 & 11.38$\pm$0.14 \\
       189.33160 & 62.215710 & 0.91 & 22.47 & 54.30 & 0.00 & 0.86 & 3.64$\pm$0.23 & 10.41$\pm$0.04 & -20.66$\pm$-0.13 & -24.62$\pm$-0.13 & 129.87$\pm$69.84 & 34.05 & 10.03$\pm$0.63 & 10.36$\pm$0.63 \\
       189.40388 & 62.242610 & 0.63 & 21.39 & 66.00 & 0.09 & 0.93 & 8.31$\pm$0.21 & 9.95$\pm$0.18 & -20.64$\pm$-0.57 & -23.49$\pm$-0.65 & 99.18$\pm$6.13 & 13.82 & 10.15$\pm$0.04 & 10.39$\pm$0.04 \\
       189.44249 & 62.244660 & 0.64 & 21.49 & 72.30 & 0.04 & 0.84 & 3.33$\pm$0.21 & 10.31$\pm$0.06 & -20.55$\pm$-0.12 & -24.38$\pm$-0.15 & 175.47$\pm$14.38 & 33.86 & 10.25$\pm$0.06 & 10.53$\pm$0.06 \\
       189.30710 & 62.253220 & 0.52 & 21.03 & 43.80 & 0.00 & 0.88 & 4.24$\pm$0.19 & 10.29$\pm$0.06 & -20.29$\pm$-0.18 & -24.13$\pm$-0.21 & 181.54$\pm$29.76 & 28.62 & 10.39$\pm$0.12 & 10.64$\pm$0.12 \\
       189.33580 & 62.274980 & 0.84 & 21.59 & 72.20 & 0.03 & 0.88 & 8.22$\pm$0.23 & 10.05$\pm$0.18 & -21.28$\pm$-0.58 & -24.09$\pm$-0.65 & 168.61$\pm$4.61 & 18.26 & 10.61$\pm$0.02 & 10.82$\pm$0.02 \\
       189.21300 & 62.175340 & 0.41 & 19.94 & 36.10 & 0.18 & 0.56 & 4.04$\pm$0.16 & 10.59$\pm$0.06 & -20.46$\pm$-0.18 & -24.70$\pm$-0.22 & 170.26$\pm$10.68 & 27.67 & 10.31$\pm$0.04 & 10.57$\pm$0.04 \\
       189.17271 & 62.181010 & 0.94 & 22.47 & 73.50 & 0.12 & 0.96 & 5.48$\pm$0.24 & 10.34$\pm$0.06 & -20.65$\pm$-0.19 & -23.74$\pm$-0.20 & 171.56$\pm$43.62 & 20.06 & 10.45$\pm$0.20 & 10.67$\pm$0.20 \\
       189.22171 & 62.188090 & 0.94 & 21.27 & 73.40 & 0.22 & 0.64 & 7.96$\pm$0.24 & 10.86$\pm$0.05 & -21.72$\pm$-0.16 & -25.64$\pm$-0.19 & 228.84$\pm$34.51 & 21.64 & 10.86$\pm$0.11 & 11.07$\pm$0.11 \\
       189.29829 & 62.190790 & 0.41 & 21.81 & 26.20 & 0.06 & 0.84 & 5.63$\pm$0.37 & 9.30$\pm$0.10 & -19.01$\pm$-0.33 & -21.57$\pm$-0.37 & 86.23$\pm$4.70 & 20.20 & 9.86$\pm$0.05 & 10.17$\pm$0.05 \\
       189.27409 & 62.257080 & 0.50 & 21.86 & 58.90 & 0.04 & 0.76 & 5.83$\pm$0.18 & 9.59$\pm$0.09 & -19.51$\pm$-0.28 & -22.40$\pm$-0.32 & 116.64$\pm$17.25 & 23.41 & 10.14$\pm$0.10 & 10.42$\pm$0.10 \\
       189.06029 & 62.121890 & 0.97 & 21.25 & 79.70 & 0.15 & 0.50 & 10.87$\pm$0.24 & 11.05$\pm$0.03 & -22.24$\pm$-0.11 & -25.20$\pm$-0.12 & 236.10$\pm$9.21 & 14.03 & 11.02$\pm$0.03 & 11.20$\pm$0.03 \\
       189.15370 & 62.126510 & 1.00 & 22.10 & 79.00 & 0.00 & 0.61 & 6.60$\pm$0.24 & 10.20$\pm$0.19 & -21.90$\pm$-0.64 & -24.80$\pm$-0.73 & 169.74$\pm$65.96 & 26.13 & 10.52$\pm$0.35 & 10.77$\pm$0.35 \\
       189.05609 & 62.153080 & 0.41 & 20.59 & 64.90 & 0.02 & 0.80 & 5.99$\pm$0.20 & 10.01$\pm$0.06 & -20.21$\pm$-0.20 & -23.20$\pm$-0.23 & 120.14$\pm$8.35 & 19.79 & 10.18$\pm$0.05 & 10.44$\pm$0.05 \\
       189.02771 & 62.164350 & 1.20 & 21.07 & 45.20 & 0.02 & 0.99 & 24.31$\pm$0.25 & 11.06$\pm$0.03 & -20.66$\pm$-0.11 & -25.25$\pm$-0.11 & 219.27$\pm$39.61 & 1.78 & 11.31$\pm$0.13 & 11.44$\pm$0.13 \\
       189.11980 & 62.173260 & 0.94 & 21.88 & 37.70 & 0.38 & 0.82 & 17.76$\pm$0.73 & 10.66$\pm$0.05 & -21.27$\pm$-0.17 & -24.44$\pm$-0.20 & 167.54$\pm$8.68 & 9.51 & 10.94$\pm$0.04 & 11.11$\pm$0.04 \\
       189.15421 & 62.199970 & 0.78 & 21.38 & 79.50 & 0.06 & 0.65 & 9.16$\pm$0.22 & 10.12$\pm$0.17 & -21.24$\pm$-0.55 & -24.09$\pm$-0.62 & 141.64$\pm$12.20 & 17.87 & 10.51$\pm$0.06 & 10.73$\pm$0.06 \\    
[1ex]   
\hline
\end{tabular}
\end{table}
\end{tiny}
\clearpage
\end{turnpage}

\end{document}